\documentclass[12pt]{article}
\usepackage[T1]{fontenc}
\usepackage[utf8]{inputenc}

\usepackage[letterpaper, portrait, margin=1in]{geometry}

\usepackage{enumitem}

\usepackage{graphicx}
\usepackage{wrapfig}

\usepackage[bookmarks]{hyperref}
\hypersetup{
  colorlinks=true,
  linkcolor=blue,
  filecolor=magenta,
  urlcolor=blue,
  citecolor=blue
}

\usepackage{amsmath}
\usepackage{amsfonts}
\usepackage{amssymb}
\usepackage{amsthm}

\usepackage[nameinlink]{cleveref}

\usepackage[ruled,vlined]{algorithm2e}

\newcommand{\indep}{\raisebox{0.05em}{\rotatebox[origin=c]{90}{$\models$}}}

\allowdisplaybreaks

\numberwithin{equation}{section}

\theoremstyle{definition}
\newtheorem*{definition*}{Definition}
\newtheorem*{intuition*}{Intuition}
\newtheorem*{assumption*}{Assumption}
\newtheorem*{remark*}{Remark}
\newtheorem*{example*}{Example}
\newtheorem*{exercise*}{Exercise}

\theoremstyle{plain}
\newtheorem*{theorem*}{Theorem}
\newtheorem*{proposition*}{Proposition}
\newtheorem*{claim*}{Claim}

\theoremstyle{definition}
\newtheorem{definition}{Definition}[section]
\newtheorem{assumption}{Assumption}[section]
\crefname{assumption}{assumption}{assumptions}
\newtheorem{remark}{Remark}[section]

\theoremstyle{plain}
\newtheorem{theorem}{Theorem}[section]
\newtheorem{proposition}[theorem]{Proposition}

\newtheorem{corollary}[theorem]{Corollary}

\usepackage{bm}
\usepackage{multirow}
\usepackage{booktabs}
\usepackage{float}

\usepackage{setspace}
\usepackage[symbol]{footmisc}
\renewcommand{\thefootnote}{\fnsymbol{footnote}}

\usepackage{natbib}

\title{On the Performance of the Neyman Allocation with Small Pilots}

\author{
Yong Cai
\footnote{Becker Friedman Institute, University of Chicago, Chicago, IL
60615, USA},
\and
Ahnaf Rafi\footnote{Department of Economics, Northwestern University, Evanston,
IL 60208, USA}
\footnote{Corresponding author. Email address:
\href{mailto: ahnafrafi@u.northwestern.edu}{ahnafrafi@u.northwestern.edu}}
}

\begin{document}

\maketitle

\begin{abstract}
The Neyman Allocation is used in many papers on experimental design, which
typically assume that researchers have access to large pilot studies. This may
be unrealistic. To understand the properties of the Neyman Allocation with small
pilots, we study its behavior in an asymptotic framework that takes pilot size
to be fixed even as the size of the main wave tends to infinity. Our analysis
shows that the Neyman Allocation can lead to estimates of the ATE with higher
asymptotic variance than with (non-adaptive) balanced randomization. In
particular, this happens when the outcome variable is relatively homoskedastic
with respect to treatment status or when it exhibits high kurtosis. We provide a
series of empirical examples showing that such situations can arise in
practice. Our results suggest that researchers with small pilots should not use
the Neyman Allocation if they believe that outcomes are homoskedastic or
heavy-tailed. Finally, we examine some potential methods for improving the
finite sample performance of the FNA via simulations.
\end{abstract}

\noindent%
{\it JEL Classification:} C21, C90

\noindent%
{\it Keywords:} Experiment design, treatment effect estimation, alternative
asymptotics
\vfill


\renewcommand{\thefootnote}{\arabic{footnote}}

\section{Introduction}

A growing literature on experiment design provides researchers with tools for
reducing the asymptotic variance of their average treatment effect (ATE)
estimates. Many do so in the context of two-wave experiments, where the
researcher has access to a pilot study that can be used to improve the main
study. Pilots are typically assumed to be large, allowing population parameters
to be well-estimated. However, large pilots may not be realistic in practice. In
this paper, we study the implications of small pilots for experiment design
through the lens of the Neyman Allocation.

The Neyman Allocation (\citealt{neyman1934two}) is a simple method for
minimizing the variance of the difference-in-means estimator of ATE. In a
setting without covariates, suppose that the standard deviations of the treated
and control outcomes are known. The Neyman Allocation assigns more units to
either treatment or control in proportion to the ratio of their standard
deviations. Intuitively, the optimal experiment entails more measurements of the
noisier quantity. Since the variances are not known in practice, the feasible
Neyman Allocation (FNA) estimates the variances using the pilot study and then
plugs the estimates into the assignment rule.

The FNA is an important part of many experiment design procedures. In the
econometrics literature alone there are several notable
works. \citet{hahn2011adaptive} propose to estimate the variance of outcome and
control groups conditional on covariate value, implementing the FNA conditional
on covariates. In a similar vein, \citet{tabord2021stratification} employs
tree-based techniques to stratify units based on their covariates. Units are
then assigned to treatment and control based on the FNA conditional on
strata. \citet{cytrynbaum2021designing} proposes local randomization
to select representative units for participation and treatment in
experiments. In what \citet{cytrynbaum2021designing} terms the ``fully
efficient" case, treatment proportion conditional on the randomization group is
chosen by the FNA. Despite their differences, the above papers study their
proposals in asymptotic frameworks that take the pilot size to infinity. Their
analyses, appropriate for large pilots, essentially assume that population
parameters are arbitrarily well-estimated from the pilots alone. In practice,
pilots are often conducted for logistical reasons and may be small. In such
settings, accurately estimating the relevant variances may be difficult. Indeed,
all of the above authors caution against the use of their methods when pilot
sizes are small.

To understand the implications of small pilots, we study the properties of the
Neyman Allocation in an asymptotic framework that takes pilot size to be fixed even as the size of the main wave tends to infinity. In this setting, uncertainty in parameter estimation is
non-negligible in the limit, and the FNA may not be the optimal allocation.  In fact, we show that the FNA can do worse than the
naive, balanced allocation that assigns half the units to treatment and half to
control. This occurs when outcomes have similar variances across treatment and
control -- that is, when outcomes are relatively homoskedastic with respect to
treatment status. To assess how much homoskedasticity exists in practice, we
examine the first 10 completed experiments in the AER RCT Registry. We ask the
hypothetical question: if researchers conducted these studies as a two-wave
experiment with a random sample from the same population, would they do better
with the FNA or with balanced randomization? We find that the treatment and
control groups are often highly homoskedastic across a range of outcomes and
across experiments. This suggests that if faced with a small pilot, the authors
of those studies would likely not have benefited from implementing the
FNA. Finally, we show that as the pilot size increases, the amount of
heteroskedasticity needed for the FNA to be preferable to the balanced
allocation decreases, but at a rate that depends on the kurtosis of the outcome
variables. Hence, even when researchers believe they are in a setting with high
heteroskedasticity, they may want to avoid the FNA if they also believe that the
outcomes are fat-tailed.

The above findings suggest that researchers should be cautious when designing
experiments using the FNA with small pilots. However, even when pilots are
large, methods which condition on many covariates may end up estimating the FNA
using a small conditional sample. Furthermore, if researchers believe that units
exhibit cluster-dependence -- a common assumption in empirical work -- the
number of ``effective" observations may be smaller still, impeding the
estimation of the FNA.

However, our results do not imply that the FNA should never be used. Instead,
researchers may consider alternative procedures with better performance under
homoskedasticity. We explore three candidate solutions via simulations. One
is based on testing for homoskedasticity, and the other two continuously
regularize the FNA towards the balanced allocation. Our simulations suggest that
the procedure we term ``exponential regularization" offers the best trade-offs
in that performance under heteroskedasticity is the least compromised in
exchange for better performance under homoskedasticity. Alternatively,
researchers might consider optimizing other aspects of the design, such as how
strata are formed, or in the selection of covariates for
stratification. \citet{bai2022optimality} and \citet{cytrynbaum2023optimal}
discuss methods to robustify these approaches for small pilots.

The challenges that small pilots pose for the FNA were noted as early as
\citet{sukhatme1935contribution}, which is concerned with both small pilots and
small main wave experiments. Given the relative intractability of the set-up,
the author concludes using simulations that the FNA performs well. Authors such
as \citet{hahn2011adaptive, tabord2021stratification, bai2022optimality} have
noted the limitations of their method in a small pilot setting. In particular,
\citet{cytrynbaum2021designing,cytrynbaum2023optimal} highlights that the FNA
leads to estimators with residual variation from variance estimation when pilots
are small. \citet{cytrynbaum2023optimal} further shows that estimating
stratification variables using small pilots may be counterproductive. In a
network setting, \citet{viviano2022experimental} characterizes the regret of the
designed experiment as a function of pilot size, providing a rule for choosing
pilot size.  Our paper builds on this body of work by setting up balanced
randomization as the natural alternative to the FNA when pilots are small,
focusing on the relative efficiency of the two methods.
Our analysis specifically addresses the use of the FNA in reducing the
asymptotic variance of the difference-in-means estimator in two-wave
experiments.
It does not speak to papers which take the treatment assignment probability as
given, such as \citet{bai2022optimality}.

Our work stands in contrast to \citet{blackwell2022batch}, who argue that
the FNA can perform well even when pilots are small.
However, their theoretical results rely on pilot sizes going to infinity, which
we consider less suitable for addressing the small pilot case.
Their simulations, which consider only normally distributed outcomes,
may understate the challenges arising from high kurtosis.

Our paper is closely related to papers pointing out a similar issue in the design
of sequential experiments. \citet{melfi1998variablility} argue by simulation
that treatment assignment rules based on estimated outcomes can do worse than
non-adaptive rules due to estimation noise. Theoretical analysis is provided in
\citet{hu2003optimality}, in an asymptotic framework that does not nest
ours. Our paper is also related to those studying small sample problems in
experiments. \citet{deChaisemartin2019level} are concerned with the problems
small strata pose for inference. \citet{bruhn2009pursuit} consider the
effectiveness of various randomization strategies in achieving balance when a
single-wave experiment is small. Finally, we note that there is a large
literature discussing the large-pilot properties of the Neyman Allocation for
alternative criteria such as power (e.g. \citealt{brittain1982optimal},
\citealt{azriel2012optimal}), minimax optimality
(e.g. \citealt{bai2021randomize}) or ethical considerations (e.g. Chapter 8 of
\citealt{hu2006theory}). These criteria fall outside the scope of this present
paper.

The remainder of this paper is organized as follows. Section
\ref{section--framework} presents the theoretical framework. Section
\ref{section--toy-example} contains analytical and simulation results using a
stylized toy example. Our main theoretical results can be found in Section
\ref{section--theoretical-results}. We assess the level of homoskedasticity in
selected empirical applications in Section
\ref{section--empirical_homosked}. Section \ref{section--conclusion} concludes
the paper. All proofs as well as additional empirical examples are contained in
the online appendix.

\section{Framework}
\label{section--framework}

We use a standard binary treatment potential outcomes framework assuming an
infinite superpopulation. The potential outcomes are \((Y (0), Y (1))\), where
\(Y (0)\) denotes the potential outcome under control or status quo and \(Y
(1)\) denotes the potential outcome under treatment or the innovation.


\begin{assumption}
\label{asm--2nd-moments}
Potential outcomes \((Y (0), Y (1)) \sim F\) have finite second moments. The vector of
means is \(\mu\) and the covariance matrix is \(\Sigma\) where
\begin{align*}
  \mu =
  & \ \mathbb{E} \left[ \left( \begin{array}{c}
    Y (0) \\
    Y (1)
  \end{array} \right) \right] = \left( \begin{array}{c}
    \mu (0) \\
    \mu (1)
  \end{array} \right) \\ \Sigma =
  & \ \mathrm{Var} \left[ \left(
    \begin{array}{c}
      Y (0) \\
      Y (1)
    \end{array} \right) \right] = \left( \begin{array}{cc}
    \sigma^{2} (0) & \rho \cdot \sigma (0) \cdot \sigma (1) \\
    \rho \cdot \sigma (0) \cdot \sigma (1) & \sigma^{2} (1)
  \end{array} \right).
\end{align*}
Additionally, assume potential outcomes have variances that are positive so that
\(\sigma^{2} (a) > 0\) for each \(a \in \{0, 1\}\).
\end{assumption}

The estimand of interest is the Average Treatment Effect (ATE), \(\theta =
\mathbb{E} [Y (1) - Y (0)]\). To estimate the ATE, the experimenter conducts a
two-wave experiment. The smaller first wave, also known as the pilot, is used
to inform the experimenter about aspects of the design of the larger main wave
(i.e. the second wave). The following assumptions about the two experimental
waves will be maintained throughout the paper. For notational clarity, the tilde
symbol (e.g. \(\widetilde{X}\)) refers to quantities associated with the pilot.

\begin{assumption}
\label{asm-1stwave-observation-process}
Potential outcomes in the pilot, denoted \(\left\{ \widetilde{Y}_{i} (0),
\widetilde{Y}_{i} (1) \right\}_{i = 1}^{m}\), consist of \(m\) i.i.d. draws from
the distribution of the random vector \((Y (0), Y (1))^{\prime}\). Treatment is
randomly assigned in the pilot so that denoting assignments by \(\left\{
\widetilde{A}_{i} \right\}_{i = 1}^{m}\),
\[
  \left\{ \widetilde{Y}_{i} (0), \widetilde{Y}_{i} (1) \right\}_{i = 1}^{m}
  \indep \left\{ \widetilde{A}_{i} \right\}_{i = 1}^{m}.
\]
\end{assumption}

\begin{assumption}
\label{asm-2ndwave-outcomes-device}
Potential outcomes in the main wave, denoted by \(\left\{ Y_{i} (0), Y_{i} (1)
\right\}_{i = 1}^{n}\), are \(n\) i.i.d. draws from the distribution of the
random vector \((Y (0), Y (1))^{\prime}\) and are independent to potential
outcomes and treatment assignments in the pilot. That is,
\[
  \left\{ Y_{i} (0), Y_{i} (1) \right\}_{i = 1}^{n} \indep \left\{
  \widetilde{Y}_{i} (0), \widetilde{Y}_{i} (1), \widetilde{A}_{i} \right\}_{i =
  1}^{m}.
\]
\end{assumption}

There are various ways in which the experimenter could implement treatment assignment in the main wave. We describe a few common ones below. Note that in all of these allocation schemes, treatment assignment is independent of potential outcomes.

\begin{itemize}
  \item \textbf{Simple Random Assignment:} For a given \(p \in (0, 1)\), let
    \[
    A^{\text{rand}}_{p, i} \overset{\text{i.i.d.}}{\sim} \text{Bernoulli}(p)~.
    \]
    The associated observed outcomes in this case are denoted \(Y_{p, i}^{\text{rand}} = Y_{i} (0) \left( 1 - A^{\text{rand}}_{p, i} \right) + Y_{i} (1) A^{\text{rand}}_{p, i}\) and the estimator for the average treatment effect is the difference-in-means estimator:
    \[
    \widehat{\theta}^{\text{rand}}_{p} = \frac{\frac{1}{n} \sum_{i = 1}^{n} Y_{p, i} A^{\text{rand}}_{p, i}}{\frac{1}{n} \sum_{i = 1}^{n} A^{\text{rand}}_{p, i}} - \frac{\frac{1}{n} \sum_{i = 1}^{n} Y_{p, i} \left( 1 - A^{\text{rand}}_{p, i} \right)}{\frac{1}{n} \sum_{i = 1}^{n} \left( 1 - A^{\text{rand}}_{p, i} \right)} ~.
    \]
  \item  \textbf{Block/Complete Randomization:} For a given \(p \in (0, 1)\),
    let $n_1 = \lfloor np \rfloor$. Randomly assign exactly $n_1$ units to
    treatment so that all $\binom{n}{n_{1}}$ assignments are equally likely.
    In other words, let \(A_{p} = (A_{p, 1}, \dots, A_{p, n})^{\prime}\) denote
    the random vector of treatment assignments.
    Then, under block/complete randomization, the distribution of
    \(A_{p}\) satisfies
    \begin{equation*}
      P(A_p = \mathbf{a}) = 1/{\binom{n}{n_{1}}} \text{ for all } \mathbf{a} \in
      \{0, 1\}^{n} \text{ such that } \sum_{i = 1}^{n} a_{i} = n_{1}.
    \end{equation*}
    Correspondingly, we observe \(Y_{p, i} = Y_{i} (0) \left( 1 - A_{p, i} \right) + Y_{i} (1) A_{p, i}\) and form the difference-in-means estimator:
    \[
    \widehat{\theta}_{p} = \frac{\frac{1}{n} \sum_{i = 1}^{n} Y_{p, i} A_{p, i}}{\frac{1}{n} \sum_{i = 1}^{n} A_{p, i}} - \frac{\frac{1}{n} \sum_{i = 1}^{n} Y_{p, i} \left( 1 - A_{p, i} \right)}{\frac{1}{n} \sum_{i = 1}^{n} \left( 1 - A_{p, i} \right)} ~.
    \]
    Balanced randomization corresponds to choosing \(p = \frac{1}{2}\). We will refer to the associated treatment assignment rule as the balanced allocation.

  \item \textbf{The Infeasible Neyman Allocation:} For any given \(p \in (0,
    1)\), it can be shown that
    \[
    \sqrt{n} \left( \widehat{\theta}_{p} - \theta \right) \overset{d}{\to} \mathcal{N} \left( 0, \frac{\sigma^{2} (1)}{p} + \frac{\sigma^{2} (0)}{1 - p} \right).
    \]
    The optimal choice of \(p\) to minimize the variance of the limiting distribution is the Neyman Allocation:
    \[
    p_{\ast} = \frac{\sigma (1)}{\sigma (1) + \sigma (0)} ~.
    \]
    The optimal treatment scheme is therefore $A_{p_{\ast}}$. The associated observed outcomes and difference-in-means estimator are denoted by
    \begin{equation*}
      \begin{split}
        Y_{p_{\ast}, i} = & \ Y_{i} (0) \left( 1 - A_{p_{\ast}, i} \right) + Y_{i} (1) A_{p_{\ast}, i} \\
        \widehat{\theta}_{p_{\ast}} = & \ \frac{\frac{1}{n} \sum_{i = 1}^{n} Y_{p_{\ast}, i} A_{p_{\ast}, i}}{\frac{1}{n} \sum_{i = 1}^{n} A_{p_{\ast}, i}} - \frac{\frac{1}{n} \sum_{i = 1}^{n} Y_{p_{\ast}, i} \left( 1 - A_{p_{\ast}, i} \right)}{\frac{1}{n} \sum_{i = 1}^{n} \left( 1 - A_{p_{\ast}, i} \right)} ~.
      \end{split}
    \end{equation*}
    Implementing the Neyman Allocation requires knowledge of the quantities \(\sigma (0), \sigma (1)\) and as such is infeasible.

  \item \textbf{The Feasible Neyman Allocation (FNA):} One feasible implementation is to use the pilot data to form a plug-in estimator for \(p_{\ast}\). We start by using pilot data to estimate potential outcome variances:
    \begin{align*}
      \widetilde{\sigma}_{m}^{2} (a) =
      & \, \frac{1}{m_{a} - 1} \sum_{i = 1}^{m} \left( \widetilde{Y}_{i}
      \mathbb{I} \left\{ \widetilde{A}_{i} = a \right\} - \frac{1}{m_{a}}
      \sum_{i = 1}^{m} \widetilde{Y}_{i} \mathbb{I} \left\{ \widetilde{A}_{i} =
      a \right\} \right)^{2} ~, \\
      \text{where } \widetilde{Y}_{i} =
      & \, \widetilde{Y}_{i} (1) \widetilde{A}_{i}
      + \widetilde{Y}_{i} (0) \left( 1 - \widetilde{A}_{i} \right) ~, \\
      \text{and } m_{a} =
      & \, \sum_{i = 1}^{m} \mathbb{I} \left\{\widetilde{A}_{i} = a \right\}~.
    \end{align*}
    The Feasible Neyman Allocation is
    \begin{equation}
      \widetilde{p} =
      \begin{cases}
        \frac{\widetilde{\sigma}_{m} (1)}{\widetilde{\sigma}_{m} (1) + \widetilde{\sigma}_{m} (0)} & \text{if } \widetilde{\sigma}_{m} (1), \widetilde{\sigma}_{m} (0) > 0 ~, \\
        \frac{1}{2} & \text{otherwise}  ~.
      \end{cases}
      \label{eqn--neyman-alloc-def}
    \end{equation}
    The latter case in \eqref{eqn--neyman-alloc-def} (where at least one \(\widetilde{\sigma}_{m} (a) = 0\)) avoids division by zero and additionally avoids the case where the entire main wave sample gets assigned to a single treatment arm. If the potential outcomes are continuously distributed, this latter case happens with zero probability. The distribution of \(\widetilde{p}\) depends on \(m\), but we omit the sample size subscript for notational convenience. The associated treatment allocation scheme is $A_{\tilde{p}}$ and the observed outcomes and difference-in-means estimator are denoted
    \begin{equation*}
      \begin{split}
        Y_{\widetilde{p}, i} = & \ Y_{i} (0) \left( 1 - A_{\widetilde{p}, i}
        \right) + Y_{i} (1) A_{\widetilde{p}, i} ~, \\
        \widehat{\theta}_{\widetilde{p}} = & \ \frac{\frac{1}{n} \sum_{i = 1}^{n} Y_{\widetilde{p}, i} A_{\widetilde{p}, i}}{\frac{1}{n} \sum_{i = 1}^{n} A_{\widetilde{p}, i}} - \frac{\frac{1}{n} \sum_{i = 1}^{n} Y_{\widetilde{p}, i} \left( 1 -  A_{\widetilde{p}, i} \right)}{\frac{1}{n} \sum_{i = 1}^{n} \left( 1 -  A_{\widetilde{p}, i} \right)} ~.
      \end{split}
    \end{equation*}
\end{itemize}

\begin{remark}
We can also define INA and FNA with simple random assignment instead of block/complete randomization. However, the latter is generally considered to be preferable to simple random assignment (see e.g. \cite{lachin1988properties}) and is the standard in fields such as Development Economics (see examples in \cite{bugni2018inference}).
\end{remark}

\section{Toy Example}
\label{section--toy-example}

In this section, we illustrate the main problems with the FNA in small samples with a toy model. In the context of this simple example, we ask the question: when does the FNA do worse than the balanced allocation? It turns out that this happens for a range of plausible values of population parameters. Section \ref{section--theoretical-results} describes the extension of our findings into more general settings.

We assume in this section that the potential outcomes are bivariate normal:
\begin{equation*}
  \begin{pmatrix}
    Y(1) \\
    Y(0)
  \end{pmatrix} \sim {N} (\mu, \Sigma)~.
\end{equation*}
Suppose we have a pilot of size $m$ where $m$ is even.  Suppose treatment is assigned deterministically as follows:
\begin{align*}
  \widetilde{A}_{i} = \begin{cases}
    1 & \mbox{ if } 1\leq i \leq \frac{m}{2} \\
    0 & \mbox{ otherwise.}
  \end{cases}
\end{align*}
Using standard arguments, it follows that the variance estimates are distributed as independent \(\chi^{2}\) random variables:
\begin{equation*}
  \widetilde{\sigma}_{m}^{2} (1) \; \indep \; \widetilde{\sigma}_{m}^{2} (0)
  \quad , \quad \left( \frac{m}{2} - 1 \right) \left( \frac{\widetilde{\sigma}_{m}^{2}
  (a)}{\sigma^{2} (a)} \right) \sim \chi^{2}_{\frac{m}{2} - 1}~.
\end{equation*}
Conditional on the pilot sample, the FNA assigns
\begin{equation*}
  \widetilde{p} = \frac{\widetilde{\sigma}_{m} (1)}{\widetilde{\sigma}_{m}
  (0) + \widetilde{\sigma}_{m} (1)}
\end{equation*}
proportion of the units in the main wave to treatment. Suppose for simplicity again
these are the first $n\widetilde{p}$ units (rounding $n\widetilde{p}$ if
necessary). Then,
\begin{equation*}
  \widehat{\theta}_{\widetilde{p}} = \left(\frac{1}{n\widetilde{p}} \sum_{i =
  1}^{n\widetilde{p}} Y_i(1) \right) - \left(\frac{1}{n - n\widetilde{p}} \sum_{i =
  n \widetilde{p}+1}^{n} Y_i(0) \right)~.
\end{equation*}
Terms in the above expression have conditional distributions:
\begin{align*}
  \left[ \frac{1}{\sqrt{n\widetilde{p}}} \sum_{i = 1}^{n\widetilde{p}} Y_i(1)
  \; \bigg\lvert \; \widetilde{p} \right] \sim
  & \, N \left( 0, \sigma^2(1) \right) ~,
  \\
  \left[\frac{1}{\sqrt{n - n\widetilde{p}}} \sum_{i = n \widetilde{p}+1}^{n}
  Y_i(0) \; \bigg\lvert \; \widetilde{p} \right] \sim
  & \, N \left( 0, \sigma^2(0) \right) ~.
\end{align*}
Furthermore, they are independent. Hence,
\begin{equation*}
  \left[ \sqrt{n}\left( \widehat{\theta}_{\widetilde{p}} - \theta \right) \; \bigg\lvert \;
  \widetilde{p} \right] \sim {N}\left(0 \; , \;
  \frac{\sigma^{2}(1)}{\widetilde{p}} + \frac{\sigma^{2} (0)}{1 -
  \widetilde{p}} \right)~.
\end{equation*}
Taking expectation over $\widetilde{p}$, we have that:
\begin{equation}\label{eqn--finite-Est-Neyman-Variance}
  \text{Var} \left[ \sqrt{n}\left( \widehat{\theta}_{\widetilde{p}} - \theta \right)
  \right] = \mathbb{E}\left[  \frac{\sigma^{2}(1)}{\widetilde{p}} +
  \frac{\sigma^{2} (0)}{1 - \widetilde{p}} \right] ~.
\end{equation}
Letting $\widetilde{p}$ be constant at $p$, we obtain the variance of the
difference-in-means estimator under simple random assignment as a special case:
\begin{equation*}
  \text{Var} \left[ \sqrt{n}\left( \widehat{\theta}_{p} - \theta \right)  \right] = \frac{\sigma^{2}(1)}{p} + \frac{\sigma^{2} (0)}{1 - p}~.
\end{equation*}
Our goal is then to compare the two expressions above when we set $p =
\frac{1}{2}$. First, note that we can rewrite
\eqref{eqn--finite-Est-Neyman-Variance} as
\begin{align*}
  \mathbb{E} \left[\frac{\sigma^{2}
    (1)}{\widetilde{p}} + \frac{\sigma^{2} (0)}{1 - \widetilde{p}} \right]
  & = \mathbb{E} \left[ \left(1 + \frac{1}{Z_m}\frac{\sigma(0)}{\sigma(1)} \right) \sigma^2(1) + \left( 1 + Z_m\frac{\sigma(1)}{\sigma(0)}\right)\sigma^2(0)  \right]~,
\end{align*}
where $Z_m \sim \sqrt{F\left(\frac{m}{2}-1, \frac{m}{2}-1\right)}$.

The Neyman Allocation does worse than balanced randomization whenever the following obtains:
\begin{align*}
  & \mathbb{E} \left[ \left(1 + \frac{1}{Z_m}\frac{\sigma(0)}{\sigma(1)} \right) \sigma^2(1) + \left( 1 + Z_m\frac{\sigma(1)}{\sigma(0)}\right)\sigma^2(0)  \right] \geq 2\sigma^2(1) + 2\sigma^2(0) \\
  \Leftrightarrow \; & \mathbb{E}\left[ \frac{1}{Z_m} + Z_m \right]\sigma(1)\sigma(0) \geq \sigma^2(1) + \sigma^2(0) \\
  \Leftrightarrow \; & \frac{\sigma^2(1)}{\sigma^2(0)} - \mathbb{E}\left[ \frac{1}{Z_m} + Z_m \right]\frac{\sigma(1)}{\sigma(0)} + 1 \leq 0 \\
  \Leftrightarrow \; & \frac{\sigma^2(1)}{\sigma^2(0)} - 2\mathbb{E}\left[ Z_m \right]\frac{\sigma(1)}{\sigma(0)} + 1 \leq 0~.
\end{align*}
The final implication uses the fact that $Z_m \overset{d}{=} 1/Z_m$ under
bivariate normality and balanced randomization. By the quadratic formula, the
above inequality is satisfied if and only if
\begin{equation}\label{eqn--toymodel_result}
  \frac{\sigma(1)}{\sigma(0)} \in C_m := \left[ \; \mathbb{E}\left[ Z_m \right] - \sqrt{\mathbb{E}\left[ Z_m \right]^2 - 1 } \; , \; \mathbb{E}\left[ Z_m \right] + \sqrt{\mathbb{E}\left[ Z_m \right]^2 - 1} \; \right] ~.
\end{equation}
Note that reciprocal symmetry together with Jensen's inequality guarantees that the discriminant is strictly positive:
\begin{equation*}
  \mathbb{E}[Z_m]^2 = \mathbb{E}[Z_m] \mathbb{E}\left[\frac{1}{Z_m}\right] > 1~.
\end{equation*}
Furthermore, we have that
\begin{equation*}
  \left( \mathbb{E}\left[ Z_m \right] + \sqrt{\mathbb{E}\left[ Z_m \right]^2 - 1} \right)\left( \mathbb{E}\left[ Z_m \right] - \sqrt{\mathbb{E}\left[ Z_m \right]^2 - 1} \right) = 1~.
\end{equation*}
In other words, the interval has the form
\begin{equation*}
    C_m = \left[\frac{1}{c_m}, c_m \right] ~,
\end{equation*}
where $c_m > 1$ is the upper bound in \eqref{eqn--toymodel_result}. Hence, there
is a range of parameter values under which the FNA does strictly
worse than balanced randomization. We first note that $1 \in C_m$ for all
$m$. This is intuitive since $p = \frac{1}{2}$ is the infeasible Neyman
Allocation when $\sigma(1)/\sigma(0) = 1$. Secondly, $x \in C_m \Leftrightarrow
1/x \in C_m$. That is, the relative performance of the FNA to the balanced
allocation does not change when we relabel treatment and control.

\subsection*{Simulation Evidence}

Given an underlying distribution, it is simple to compute $C_m$ by Monte Carlo
integration. In this subsection, we present the values of $C_m$ for some simple
models and argue that for plausible values of $\sigma(1)/\sigma(0)$, the FNA
performs worse than the balanced allocation.

\begin{figure}[htpb]
\centering
\includegraphics[width=0.8\linewidth]{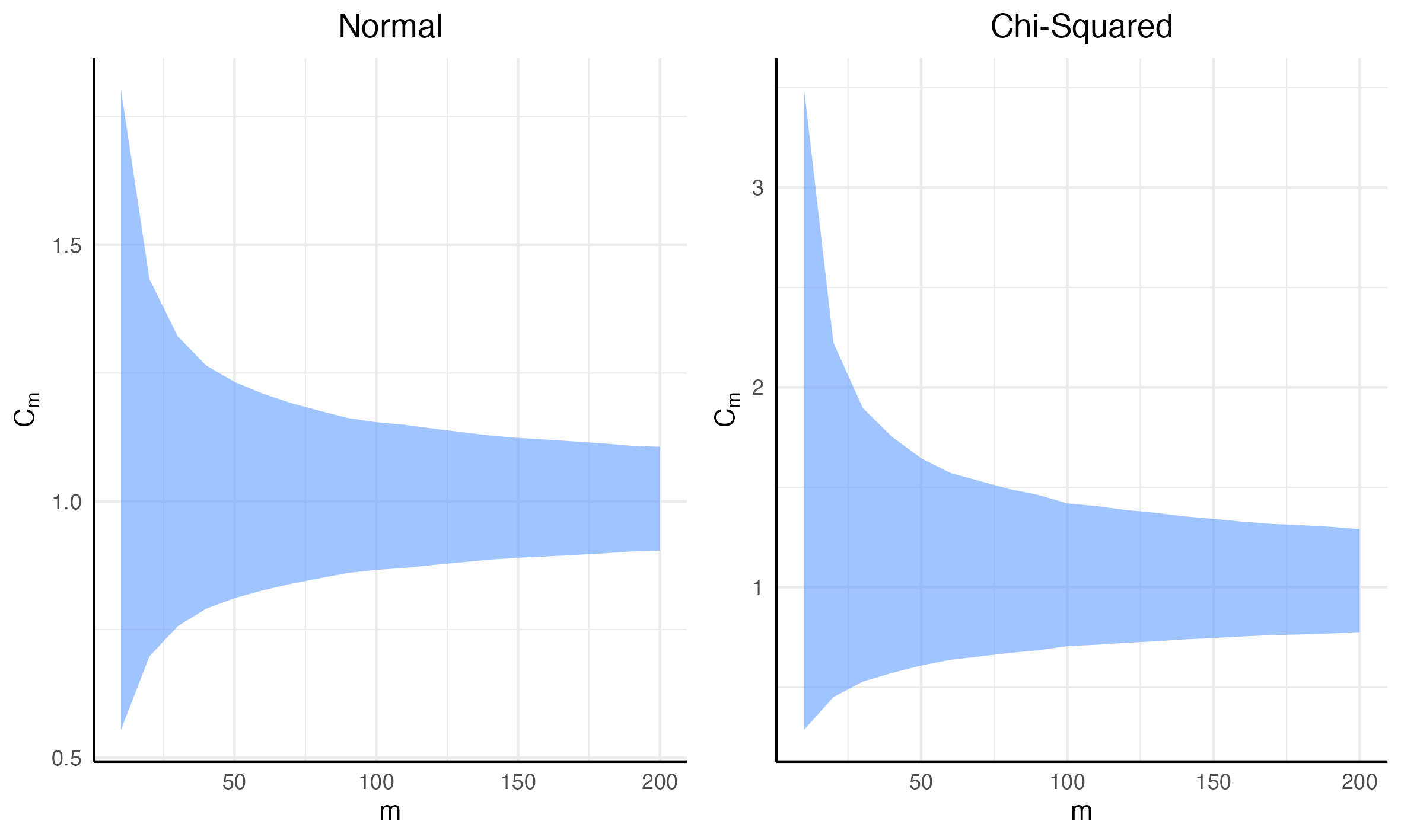}
\caption{$C_m$ when $Y(a)\sim \mu(a) + \sigma(a) \cdot N(0,1)$ and $Y(a)\sim \mu(a) + \sigma(a) \cdot \chi^2_1$ for $a \in \{0,1\}$. Monte Carlo integration using 10,000 draws.}
\label{fig:cm_normalchisq}
\end{figure}

We start with the toy model, where $Y(1) \sim {N}(\mu(1), \sigma^2(1))$
and $Y(0) \sim {N}(\mu(0), \sigma^2(0))$. Results are shown in the left panel of Figure
\ref{fig:cm_normalchisq}. The set of parameter values over which the FNA does worse is colored in blue.
This area is larger when $m$ is smaller. In fact, we show in Section
\ref{section--theoretical-results} that if the data generating process (DGP) is
sub-Gaussian, the length of $C_m$ is $O\left(m^{-1/2}\right)$. In the toy
model, $C_{20} = [0.70, 1.43]$, while $C_{50}$ is $[0.81, 1.23]$. Suppose instead that $Y(1)\sim \mu(1) + \sigma(1) \chi^2_1$, $Y(0)\sim \mu(0) +
\sigma(0) \chi^2_1$. The right panel of Figure \ref{fig:cm_normalchisq} shows that \(C_{m}\) is wider
across the range of \(m\). In particular, $C_{20} = [0.49,
2.22]$, while $C_{50} = [0.64, 1.61]$. While the
intervals may appear rather narrow at first glance, we provide numerous examples
in Section \ref{section--empirical_homosked} in which the amount of
heteroskedasticity falls within this range.

\begin{figure}[htpb]
\centering
\includegraphics[width=1\linewidth]{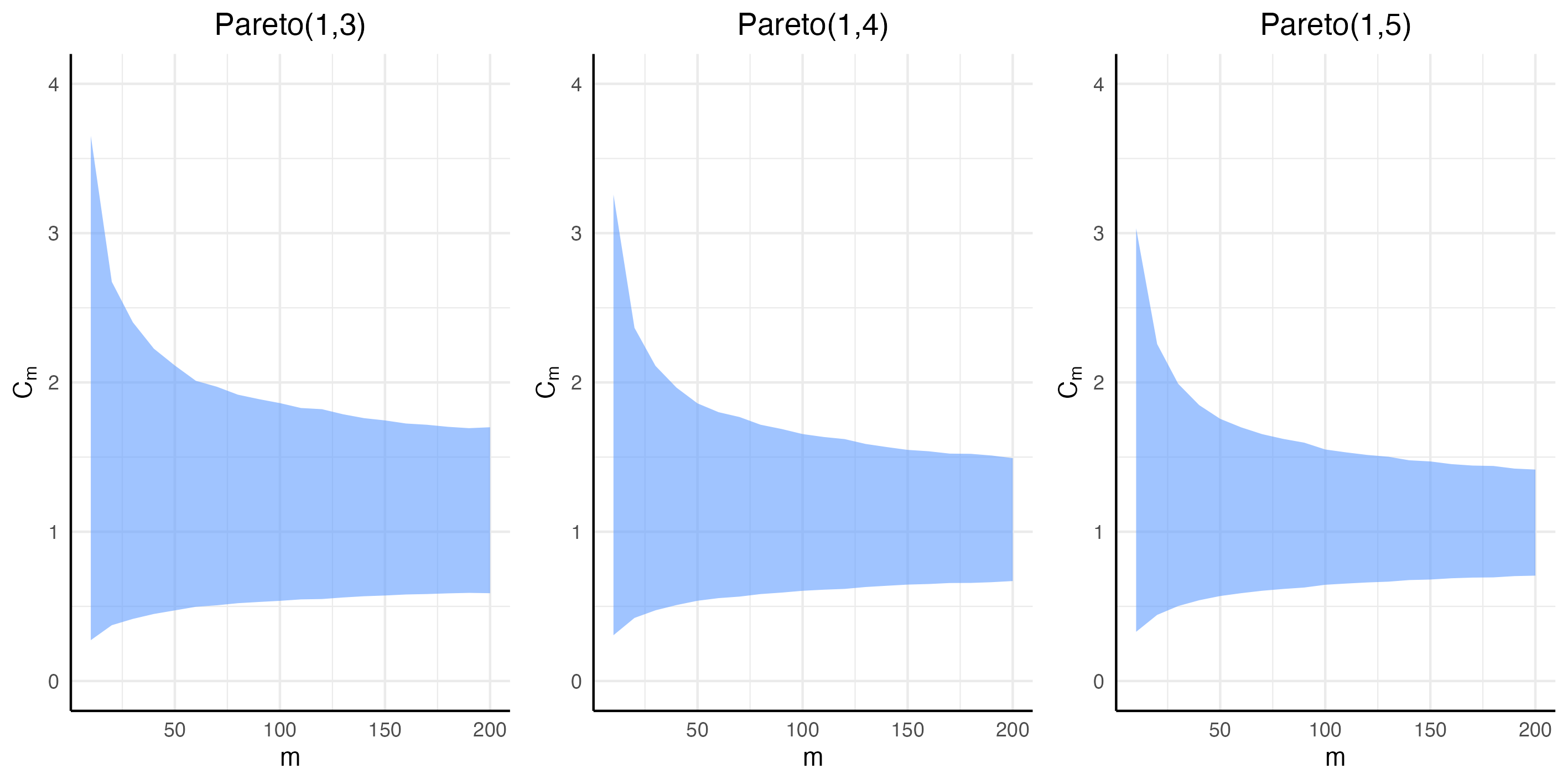}
\vspace{-5mm}
\caption{$C_m$ when $Y(a)\sim \mu(a) + \sigma(a) \cdot \text{Pareto}(1, s)$, for $a \in \{0,1\}$ and $s \in \{3,4,5\}$. Monte Carlo integration using 10,000 draws.}
\label{fig:cm_pareto}
\end{figure}

As our toy model shows, estimation of $\widetilde{\sigma}(1)$ and
$\widetilde{\sigma}(0)$ causes problems when $m$ is small. It is therefore
intuitive that $C_m$ will be wide when the distributions are fat tailed, that
is, when kurtosis is high. As such, we consider the following parametrization:
$Y(1)\sim \mu(1) + \sigma(1) \cdot \text{Pareto}(l, s)$ and $Y(0)\sim \mu(0) +
\sigma(0) \cdot \text{Pareto}(l, s)$. Here, $l$ and $s$ are the location and
scale parameters respectively. Figure \ref{fig:cm_pareto} plots $C_m$ for $l =
1$ and $s \in \{3,4,5\}$. Indeed, we see that the bands are much wider than in
the previous models. For $s = 3$, even when $m = 50$, $C_m = [0.50, 2.15]$.
Given the examples in Section \ref{section--empirical_homosked}, this appears to
be fairly extreme amounts of heteroskedasticity. Finally, we note that, $C_m$
decreases in width as we move from $s = 3$ to $s = 5$.

In sum, we see that the FNA does worse than the balanced allocation when
the treatment and control groups are relatively homoskedastic. Furthermore,
\(C_{m}\) can be quite large when \(m\) is small. The small pilot problem is
exacerbated when observations exhibit cluster dependence, so that the
``effective observations'' are fewer in number. Small pilot issues can also
arise when researchers perform stratified randomization with many strata, so
that each stratum ends up with few observations. In Section
\ref{section--empirical_homosked}, we argue that many empirical applications in
fact have fairly homoskedastic outcomes. As such, unless a researcher has reason
to believe that their outcomes are highly heteroskedastic, they should exercise
caution in using the FNA with small pilots.

\section{Theoretical Results}
\label{section--theoretical-results}

In this section, we study the theoretical properties of the FNA. We first review known results on properties of the FNA under large and small pilot asymptotics. We then present novel results in which we compare the efficiency of FNA to balanced randomization.

\subsection{Review on Properties of the FNA}

\subsubsection*{Large-\(m\) asymptotics }

Consider an asymptotic regime where both \(m, n \to \infty\), corresponding to situations where both the pilot and the main wave are large. It is straightforward to see that $\widetilde{p} \overset{p}{\to} p_*$. Furthermore, we have that:

\begin{proposition}[\cite{hahn2011adaptive}, Theorem 1]
Under \Cref{asm--2nd-moments,asm-1stwave-observation-process,asm-2ndwave-outcomes-device}, as \(m, n \to \infty\)
\[
  \sqrt{n} \left( \widehat{\theta}_{\widetilde{p}} - \theta \right)
  \overset{d}{\to} \mathcal{N} \left( 0, \Sigma_{\ast} \right),
\]
where
\[
  \Sigma_{\ast} = \frac{\sigma^{2} (1)}{p_{\ast}} + \frac{\sigma^{2} (0)}{1 -
  p_{\ast}} = (\sigma (1) + \sigma (0))^{2}.
\]
\end{proposition}

The above result obtains by applying Theorem 1 in \citet{hahn2011adaptive} to the case with only one stratum. In words, \( \widehat{\theta}_{\widetilde{p}} \) and \(
\widehat{\theta}_{p_{\ast}} \) have the same limiting distribution after
suitable centering and scaling. This occurs because in the limit, noise coming
from estimating $p_*$ with \(\widetilde{p}\) is negligible in comparison to
the sampling error of the difference-in-means estimator. Researchers employing
the large $m$ framework essentially assume that \(\widetilde{p}\) is an
arbitrarily good estimator for $p_*$ so that its sampling error can be
ignored. In practice, error in \(\widetilde{p}\) can be large, particularly
when $m$ is small. Asymptotic approximations that do not take this into account
will likely perform poorly in finite sample.

\begin{remark}
We note two minor differences in the set-up of \citet{hahn2011adaptive}: (1) Their difference-in-means estimator uses both main wave and pilot observations. The main wave allocation therefore needs to be adjusted since half of the pilot is treated. (2) They assume that $m$ is of the same order as $n$. These differences do not affect the result.
\end{remark}

\subsubsection*{Fixed-\(m\) Asymptotics}

To better understand the behavior of the FNA under small pilots, we follow \citet{cytrynbaum2021designing} in studying its
properties in an asymptotic framework that takes $m$ to be fixed, even as $n \to \infty$.

A growing literature in econometrics uses fixed-sample asymptotics to study settings in which the ``effective sample size" is small. For example, when data exhibit cluster-dependence, settings with few clusters pose unique challenges for estimation and inference. To tailor their analyses to these problems, papers such as \citet{ibragimov2010t}, \citet{canay2017randomization} and \citet{canay2021wild}  employ asymptotic frameworks in which the number of clusters is fixed in the limit. Similarly, to model difference-in-differences studies involving few treated units, \citet{conley2011inference} keep the number of treated units fixed even as the number of untreated units tend to infinity. In the same vein, inference and specification testing for regression discontinuity designs typically involves few observations around the discontinuity. To capture this, \citet{2017canayApproximatePermutationTests} analyze a permutation test for continuity in the distribution of baseline covariates under an asymptotic regime with a fixed number of observations on either side of the discontinuity.

The small pilot approach is similar in spirit to these papers. In keeping $m$
fixed, $\widetilde{p}$ is a noisy estimate of $p_*$ even in the
limit. Preserving this important feature of the statistical problem makes our
chosen framework more appropriate for analyzing experiments with small
pilots. In this setting, $\widehat{\theta}_{\widetilde{p}}$ converges in
distribution to a mixture of Gaussians instead of $\mathcal{N}(0,
\Sigma_*)$. Specifically, the form of the limiting mixture distribution depends
on the distribution of $\widetilde{p}$. This is the content of the following
result due to \citet{cytrynbaum2021designing} (also see the discussion following
equation (3.5) of \citet{cytrynbaum2023optimal}):

\begin{proposition}[\cite{cytrynbaum2021designing}, Theorem 3.17]
\label{prop-small-m-limit-dist-is-mixture}
Under \Cref{asm--2nd-moments,asm-1stwave-observation-process,asm-2ndwave-outcomes-device}, if \(m\) remains fixed as \(n \to \infty\),
\[
  \sqrt{n} \left( \widehat{\theta}_{\widetilde{p}} - \theta \right) \overset{d}{\to} \mathcal{L}_{m}~,
\]
where \(\mathcal{L}_{m}\) is a random variable whose distribution takes the form
\[
  \mathbb{P} \left( \mathcal{L}_{m} \leq t \right) = \int_{0}^{1} \Phi \left( \frac{t}{s \left( p \right)} \right) \; G_{m} (\mathrm{d} p),
\]
\(\Phi\) is the CDF of \({N} (0, 1)\), \(G_{m}\) is the distribution of \(\widetilde{p}\) and \(s (\cdot)\) is defined by
\[
  s (p) = \sqrt{\frac{\sigma^{2} (1)}{p} + \frac{\sigma^{2} (0)}{1 - p}} \ .
\]
\end{proposition}

\begin{remark}
The above proposition can be obtained from Theorem 3.17 of
\citet{cytrynbaum2021designing} by specializing it to the case with unstratified
sampling and assignment ($q = 1, \psi = 1$).
Conditional normality then implies a mixed normal limiting distribution by a
dominated convergence argument. Also see the related intuition below.
\end{remark}

\begin{remark}
Earlier versions of this paper present the same mixed normality result but
derived under simple random assignment, which is not directly implied by
\citet{cytrynbaum2021designing}.
\end{remark}

\begin{remark}
The weak limit \(\mathcal{L}_{m}\) in \Cref{prop-small-m-limit-dist-is-mixture} has mean zero. To see this, note that conditional on a value of the assignment probability \(p\), i.e. conditional on the event \(\widetilde{p} = p\), \(\mathcal{L}_{m}\) has a \({N} (0, s (p))\) distribution (where \(s (p)\) is as defined in \Cref{prop-small-m-limit-dist-is-mixture}). The conclusion then holds by the Law of Iterated Expectations.
\end{remark}

To see the intuition for this result, recall that for each \(p \in (0, 1)\),
\[
  \sqrt{n} \left( \widehat{\theta}_{p} - \theta \right) \overset{d}{\to} {N} \left( 0, \frac{\sigma^{2} (1)}{p} + \frac{\sigma^{2} (0)}{1 - p} \right) = {N} \left( 0, s^{2} (p) \right).
\]
When \(m\) is held fixed, in the limit as \(n \to \infty\), \(\widetilde{p}\) remains a non-degenerate random variable. In particular, the limiting distribution of $\widetilde{p}$ is its finite-sample distribution. Thus, the distribution of \(\sqrt{n} \left( \widehat{\theta}_{\widetilde{p}} - \theta \right)\) becomes a mixture of the marginal distributions of the process
\[
  \left\{ \sqrt{n} \left( \widehat{\theta}_{p} - \theta \right) : p \in (0, 1) \right\}~,
\]
where the mixing distribution is the distribution of \(\widetilde{p}\), denoted here by \(G_{m}\).

\subsection{When is the FNA preferred?}
\label{section--implications}

Proposition \ref{prop-small-m-limit-dist-is-mixture} shows that when pilots are small, $\hat{\theta}_{\tilde{p}}$ has asymptotic distribution that is different from $\hat{\theta}_{p_\ast}$. It stands to reason that the FNA may not be optimal in this regime. In this sub-section, we compare the asymptotic variances of $\widehat{\theta}_{\widetilde{p}}$, \(\widehat{\theta}_{p_{\ast}}\) as well as \(\widehat{\theta}_{p}\) with \(p = \frac{1}{2}\).
We first show that \(\widehat{\theta}_{\widetilde{p}}\) has larger asymptotic variance in the fixed-$m$ regime than in the large-$m$ regime. We next show that under reasonable ranges of parameter values, the asymptotic variance of \(\widehat{\theta}_{\widetilde{p}}\) can exceed that of \(\widehat{\theta}_{p}\) with \(p = \frac{1}{2}\).

\begin{remark}
It can be shown that \(\widehat{\theta}_{\widetilde{p}}\),
\(\widehat{\theta}_{p_{\ast}}\) and \(\widehat{\theta}_{p}\) are all unbiased
estimators of \(\theta\) in finite samples.
Therefore, comparing asymptotic mean squared errors is equivalent to comparing
the variances of the limit distributions.
\end{remark}

We begin with the following observation:
\begin{corollary}\label{corollary--suboptimal_variance}
    Under \Cref{asm--2nd-moments,asm-1stwave-observation-process,asm-2ndwave-outcomes-device}, suppose \(m\) remains fixed as $n \to \infty$. $\mathcal{L}_m$ has variance:
  \begin{equation*}
    \mathbb{E} \left[\frac{\sigma^{2}(1)}{\widetilde{p}} + \frac{\sigma^{2} (0)}{1 - \widetilde{p}} \right] > \Sigma_*~.
  \end{equation*}
\end{corollary}
In words, the asymptotic variance of $\widehat{\theta}_{\widetilde{p}}$ is larger under the fixed-$m$ regime than under the large-$m$ regime. When pilots are small, uncertainty in $\widetilde{p}$ may be large and could affect the asymptotic variance of $\widehat{\theta}_{\widetilde{p}}$. In particular, $\widehat{\theta}_{\widetilde{p}}$ will not be able to attain the optimal asymptotic variance of the infeasible allocation $p_*$. Conventional large-$m$ asymptotics may be too optimistic about the effectiveness of the Neyman Allocation with small pilots.

In addition to not attaining $\Sigma_*$, the $\widehat{\theta}_{\widetilde{p}}$ can do worse than $\widehat{\theta}_p$ for certain values of $\sigma^2(1)$ and $\sigma^2(0)$, as our next two results asserts. For convenience, define the following:
\begin{definition}
  Let $C_m$ be the set such that for a pilot of size $m$, $\widehat{\theta}_{\widetilde{p}}$ has higher asymptotic variance than $\widehat{\theta}_p$ if and only if $\frac{\sigma(1)}{\sigma(0)} \in C_m$. Let $|C_m|$ denote the length of $C_m$.
\end{definition}
\begin{definition} Let
\begin{equation*}
  Z_m = \frac{\widetilde{\sigma}_m(1)}{\sigma(1)} \bigg/ \frac{\widetilde{\sigma}_m(0)}{\sigma(0)}~.
\end{equation*}
and
\begin{equation*}
  B_m = \frac{1}{2} \, \mathbb{E}\left[ \frac{1}{Z_m} + Z_m \right]~.
\end{equation*}
\end{definition}
Here, $Z_m$ is our estimator of the ratio of $\sigma(1)/\sigma(0)$, appropriately normalized. $B_m-1$ is therefore the average bias of the estimators for $\sigma(1)/\sigma(0)$ and $\sigma(0)/\sigma(1)$, up to normalization. Our next proposition characterizes the region $C_m$ in terms of $B_m$:
\begin{proposition}\label{theorem--Cm_small_m}
  Under \Cref{asm--2nd-moments,asm-1stwave-observation-process,asm-2ndwave-outcomes-device}, suppose \(m\) remains fixed as $n \to \infty$.  Then
  \begin{equation*}
    C_m = \left[\frac{1}{c_m}\, , \, c_m\right]~.
  \end{equation*}
  Furthermore,
  \begin{equation*}
    c_m = B_m + \sqrt{B_m^2 - 1} > 1 \quad \text{and} \quad   \left\lvert C_m \right\rvert = 2\sqrt{B_m^2 - 1} \; > \; 0~.
  \end{equation*}
\end{proposition}

The properties of $C_m$ are intuitive. Firstly, $x \in C_m$ implies that $1/x \in C_m$. That is, the relative performance of the FNA to the balanced allocation does not change when we relabel treatment and control.
Secondly, $1 \in C_m$. This is because when $ {\sigma(1)}/{\sigma(0)} = 1$, the balanced allocation is optimal. Finally, note that $|C_m|$ depends on the bias of $\widetilde{\sigma}_m(1)/\widetilde{\sigma}_m(0)$ and $\widetilde{\sigma}_m(0)/\widetilde{\sigma}_m(1)$. In particular, if both terms are unbiased, $B_m = 1$ and $|C_m| = 0$. However, $\left\lvert C_m \right\rvert$ is strictly positive as long as $\widetilde{p}$ is not degenerate.

The exact properties of $C_m$ depend on the underlying distributions of
potential outcomes. To understand its behavior in a more general setting, we
study its first-order approximation under a sub-Gaussian assumption.
Recall that \(F\) is the distribution of \((Y (0), Y (1))\) in Assumption
\ref{asm--2nd-moments}.

\begin{definition}[Sub-Gaussian]
  Let $Y$ be a random variable. We say that $Y$ is sub-Gaussian if there exists $K$ such that
  \begin{equation*}
    \mathbb{P}\left(|X| \geq t\right) \leq 2 \exp\left(-t^2/K^2\right) \quad \mbox{ for all } t \geq 0~.
  \end{equation*}
  We say that $F$ is sub-Gaussian if $Y(1)$ and $Y(0)$ are sub-Gaussian.
\end{definition}
That is, a random variable is sub-Gaussian if its tails decay at least as quickly as a normal/Gaussian distribution.
This yields the following:
\begin{proposition}\label{theorem--Cm_large_m}
Under
\Cref{asm--2nd-moments,asm-1stwave-observation-process,asm-2ndwave-outcomes-device},
suppose $n \to \infty$. Suppose additionally that $F$ is sub-Gaussian. Then,
\begin{equation*}
  C_m = \left[ 1 - \sqrt{\frac{V}{m}} + \delta_m^- \, , \,
  1 + \sqrt{\frac{V}{m}} + \delta_m^+ \right]~,
\end{equation*}
where $\delta_m^+, \delta_m^-  = o\left(\frac{1}{\sqrt{m}}\right)$ and
\begin{gather*}
  V  = \frac{1}{4} \left(\frac{\mathbb{E}\left[ \left( Y_i(1) - \mu(1)
  \right)^4 \right]}{\sigma^4(1)} + \frac{\mathbb{E}\left[ \left( Y_i(0) -
  \mu(0) \right)^4 \right]}{\sigma^4(0)} - 2 \right)~.
\end{gather*}
\end{proposition}

Provided that the potential outcomes are sub-Gaussian, the relative efficiency
of $\widehat{\theta}_{\widetilde{p}}$ and $\widehat{\theta}_p$ under $p = 1/2$
is, to a first order, determined by the kurtosis of $Y_i(1)$ and
$Y_i(0)$. Intuitively, if the potential outcomes have fatter tails,
$\widetilde{p}$ will be poorly estimated, leading to larger variance in
$\widehat{\theta}_{\widetilde{p}}$.

Furthermore, $|C_m|$ is shrinking to $0$ at the rate $1/\sqrt{m}$. Letting $m
\to \infty$, $\widehat{\theta}_{\widetilde{p}}$ has weakly lower asymptotic
variance across all parameter values. Hence, we recover the classic result
concerning the optimality of the Neyman Allocation. When $m$ is small, however,
$C_m$ can be wide, as Proposition \ref{theorem--Cm_small_m} suggests. As we will
argue in Section \ref{section--empirical_homosked}, many empirical applications
have $\sigma(1)/\sigma(0)$ close to $1$, so that for small $m$, these ratios
fall within the range in which balanced randomization is preferred.

While the sub-Gaussian assumption limits the applicability of Proposition
\ref{theorem--Cm_large_m}, it covers binary and bounded outcomes, which are
relevant in empirical work. Furthermore, we consider it to be a negative result:
even when potential outcomes are well-behaved, the FNA is sensitive to the
kurtosis of the potential outcomes. It can still perform poorly relative to the
balanced allocation as a result.

\subsection{Efficiency Loss from the FNA}

The previous section showed that the FNA can perform worse than the balanced
allocation when the pilots are small. This raises the question: what is the
efficiency loss from using the FNA in such scenarios?
To answer this question, consider the following criteria:

\begin{definition}[Efficiency Loss of FNA under $F$]\label{definition--eff_loss}
For \((Y (0), Y (1)) \sim F\), denote
\begin{align*}
\mathcal{L}^d(F) =  \mathbb{E} \left[\frac{\sigma^{2}(1)}{\widetilde{p}} + \frac{\sigma^{2} (0)}{1 - \widetilde{p}} \right] - 2\left(\sigma^2(1) + \sigma^2(0)\right)~,\\
\quad \mathcal{L}^r(F) =    \mathbb{E} \left[\frac{\sigma^{2}(1)}{\widetilde{p}} + \frac{\sigma^{2} (0)}{1 - \widetilde{p}} \right] \bigg/ 2\left(\sigma^2(1) + \sigma^2(0)\right)~.
\end{align*}
\end{definition}
In words, we measure efficiency loss using either the difference or the ratio of the asymptotic variances arising from the FNA and the balanced allocation. Larger values of $\mathcal{L}^d$ and $\mathcal{L}^r$ indicate that the FNA is doing worse relative to balanced randomization.

Clearly, the efficiency losses depend on $F$. For a given $F$, the two criteria are equivalent in the sense that the difference criterion is positive if and only if the ratio criterion exceeds one. Depending on the set-up, one will be more convenient to work with than the other. From Definition \ref{definition--eff_loss}, it is straightforward to see that:
\begin{proposition}\label{theorem--eff_loss}
  Under \Cref{asm--2nd-moments,asm-1stwave-observation-process,asm-2ndwave-outcomes-device}, suppose \(m\) remains fixed as $n \to \infty$.  Then
  \begin{align*}
    \mathcal{L}^d(F) & = 2\left(B_m - 1\right) \sigma(1)\sigma(0) - \left(\sigma(1) - \sigma(0)\right)^2~, \\
    \mathcal{L}^r(F) & = \frac{1}{2} + B_m \cdot \frac{\sigma(1)\sigma(0)}{\sigma^2(1) + \sigma^2(0)}~.
  \end{align*}
  Furthermore, if $F$ is sub-Gaussian, then
  \begin{align*}
    \mathcal{L}^d(F) & = \frac{V}{m} \cdot \sigma(1)\sigma(0) - \left(\sigma(1)
    - \sigma(0)\right)^2 + \delta^d_m~, \\
    \mathcal{L}^r(F) & = \frac{1}{2} + \left(1 + \frac{V}{2m}\right) \cdot \frac{\sigma(1)\sigma(0)}{\sigma^2(1) + \sigma^2(0)} + \delta^r_m~,
  \end{align*}
  where $\delta^d_m, \delta^r_m = o(1/m)$.
\end{proposition}
The above proposition shows us how the efficiency loss from using the FNA
depends on $B_m$ as well as the amount of heteroskedasticity in $F$. To better
understand the trade-off from the two terms, suppose we fix $B_m$ and
reparametrize $\sigma(0) = 1/\sqrt{1+\rho^2}$, $\sigma(1) =
\rho/\sqrt{1+\rho^2}$. Now let us consider the effect of changing $\rho$. This
is the thought experiment in which we change the relative scaling of
$\varepsilon(1) = Y(1) - \mu(1)$ and $\varepsilon(0) = Y(0) - \mu(0)$ while
keeping their functional forms fixed. Additionally, this entails keeping
$\text{AVar}(\hat{\theta}_{\frac{1}{2}}) = \text{Var}(Y(1)) + \text{Var}(Y(0))$
fixed at $2$. This scaling is useful for discussing the properties of
$\mathcal{L}^d$ since it scales linearly with  $\sigma(0)$ and $\sigma(1)$.
$\mathcal{L}^r$ is invariant to the scale of $\sigma(0)$ and $\sigma(1)$. With
this parametrization, we have that:
\begin{equation*}
  \frac{\partial \mathcal{L}^d}{\partial \rho} = \frac{2B_m(1-\rho^2)}{(1+\rho^2)^2} \quad \mbox{ and } \quad   \frac{\partial \mathcal{L}^r}{\partial \rho}  = \frac{B_m\left(1- \rho^2\right)}{(1+\rho^2)^2}~.
\end{equation*}
The above expressions are intuitive. When $\rho$ exceeds $1$, increasing $\rho$ amounts to increasing heteroskedasticity. FNA is more preferable to balanced randomization and the ratio loss decreases.

Furthermore, it straightforward to see that both $\mathcal{L}^d$ and $\mathcal{L}^r$ are maximized when $\rho = 1$, with the respective values of $B_m$ and $\frac{1}{2}(1+B_m)$. On the other hand, $\mathcal{L}^d$ and $\mathcal{L}^r$ have minimum values $-1$ and $\frac{1}{2}$ respectively, which are obtained when either $\rho = 0$ or $\rho \to \infty$. In summary, for a given functional form of $\varepsilon(0)$ and $\varepsilon(1)$, the loss from using FNA depends on $B_m$ but the maximum possible gain does not. This asymmetry foreshadows the next result.

Since $F$ is unknown, we proceed by comparing the supremum and infimum of $\mathcal{L}^d(F)$ and $\mathcal{L}^r(F)$ (i.e. maximum losses and gains) over appropriate classes of DGPs. To that end, consider the following.
\begin{definition}
For \(K < \infty\), let
  \begin{align*}
    \mathcal{F}(K) = \left\{ F \mbox{ s.t. } 0 <  \sigma^2 (0), \sigma^2(1) \leq K \right\},
  \end{align*}
where it should be understood that \(\sigma^{2} (0), \sigma^{2} (1)\) both
depend on \(F\).
Furthermore, let
\begin{align*}
    \mathcal{F} (\infty) = \left\{ F \mbox{ s.t. } 0 < \sigma^2 (0), \sigma^2(1)
    < \infty \right\}.
\end{align*}
\end{definition}
Then we have that:
\begin{proposition}\label{theorem--regret}
  Under \Cref{asm--2nd-moments,asm-1stwave-observation-process,asm-2ndwave-outcomes-device}, suppose \(m\) remains fixed as $n \to \infty$. Then,
  \begin{align*}
    \sup_{F \in \mathcal{F}(K)} \mathcal{L}^d(F) = \infty \quad &, \quad \inf_{F \in \mathcal{F}(K)} \mathcal{L}^d(F) = -K~, \\
    \sup_{F \in \mathcal{F}(\infty)} \mathcal{L}^r(F) = \infty \quad &, \quad \inf_{F \in \mathcal{F}(\infty)} \mathcal{L}^r(F) = \frac{1}{2}~.
  \end{align*}
\end{proposition}
In words, the worst case loss from using the FNA -- whether measured as a
difference or as a ratio relative to the variance of the balanced allocation --
is unbounded. On the other hand, the gains are bounded. Most strikingly, the
fact that $\inf_{F \in \mathcal{F}(\infty)} \mathcal{L}^r(F) = \frac{1}{2}$
suggests that the best a researcher can do with the FNA is to obtain main wave
confidence intervals for $\hat{\theta}$ that are $1/\sqrt{2} \approx 70\%$ the
length of that from balanced randomization. This comes at the cost of
potentially doing much worse as a result of noise from the pilot estimation.

We can also interpret the result in terms of regret. Suppose a researcher is
constrained to either using the FNA or the balanced allocation. If their utility
over the second stage estimator is given by $U(\hat{\theta}) =
-\left({\hat{\theta}-\theta}\right)^2$, then $\mathcal{L}^d(F)$ is the
asymptotic regret of choosing the FNA, while $-\mathcal{L}^d(F)$ is the
asymptotic regret of choosing the balanced allocation. A similar argument can be
made for $\mathcal{L}^r(F)$ with $U(\hat{\theta}) =
-\log\left({\hat{\theta}-\theta}\right)$. Proposition \ref{theorem--regret} then
says that the regret of using the FNA over the balanced allocation is unbounded,
while the regret of using the balanced allocation over the FNA is not. The
balanced allocation is therefore preferred in terms of maximum regret.
This is admittedly a very conservative way of choosing an estimation procedure.
However, it is coherent in the following sense.
Researchers may prefer the FNA if they want to guard against large amounts of
heteroskedasticity in the population.
However, if researchers are being agnostic about $\sigma(1)/\sigma(0)$, they
should also be agnostic about $F$, in which case the balanced allocation may
ultimately be preferable.

\begin{remark}
Proposition \ref{theorem--regret} continues to hold if we further restrict $F$
so that $Y(0)$ and $Y(1)$ have uniformly bounded $a^\text{th}$ moments for any
$a$.
\end{remark}

\begin{remark}
In the proof of Proposition \ref{theorem--regret}, we construct a sequence of
DGPs over which the variance of the $\hat{\theta}_{\tilde{p}}$ diverge. This
implies that $\hat{\theta}_{\tilde{p}}$ is not uniformly consistent with large
pilots.
\end{remark}

\begin{remark}
It is also known in the large pilot case that $\inf_{F \in \mathcal{F}(\infty)}
\mathcal{L}^r(F) = \frac{1}{2}$. See for example \citet{obradovic2023using} and
\citet{zhao2023adaptive}.
\end{remark}

\subsection{Simulation Evidence}\label{section--mse_sims}

The efficiency losses recorded in Proposition \ref{theorem--regret} are
pessimistic and may not be relevant in practice. To better appreciate the losses
that researchers might experience, we next present mean-squared errors for
$\hat{\theta}_{\tilde{p}}$ under various data-generating processes, computed via
Monte Carlo simulations. Our choice of parameter values is inspired by
\citet{blackwell2022batch}.

Let $Y(1) = 0.075 + \sigma(1)\varepsilon(1)$ and $Y(0) =
\sigma(0)\varepsilon(0)$. We are interested in the MSE of
$\hat{\theta}_{\tilde{p}}$ as $\sigma(1)/\sigma(0)$ varies over $[0,1]$. We do
not need to consider values greater than $1$ given the reciprocal symmetry. We set $\sigma^2(1) + \sigma^2(0) = 2$ so that the variance of $\hat{\theta}_{\frac{1}{2}}$ is fixed. As in Section \ref{section--toy-example}, consider the following models:
\begin{itemize}
  \item[] \textbf{Model 1:} $\varepsilon(1), \varepsilon(0) \sim N(0,1)$
  \item[] \textbf{Model 2:} $\varepsilon(1), \varepsilon(0) \sim (\chi^2_1 - 1)/\sqrt{2}$
  \item[] \textbf{Model 3:} $\varepsilon(1), \varepsilon(0) \sim (\text{Pareto}(1,3) - \frac{3}{2})/\sqrt{3/4}$
  \item[] \textbf{Model 4:} $\varepsilon(1), \varepsilon(0) \sim (\text{Pareto}(1,4) - \frac{4}{3})/\sqrt{2/9}$
  \item[] \textbf{Model 5:} $\varepsilon(1), \varepsilon(0) \sim (\text{Pareto}(1,5) - \frac{5}{4})/\sqrt{5/48}$
\end{itemize}
where $\varepsilon(1)$ and $\varepsilon(0)$ are standardized to have mean $0$ and variance $1$. We set the pilot to have $m = 20$, while main wave has $n = 1000$. We run the simulations with 100,000 draws for each set of parameter values.

\begin{figure}[htpb]
\centering
\includegraphics[width=0.66\linewidth]{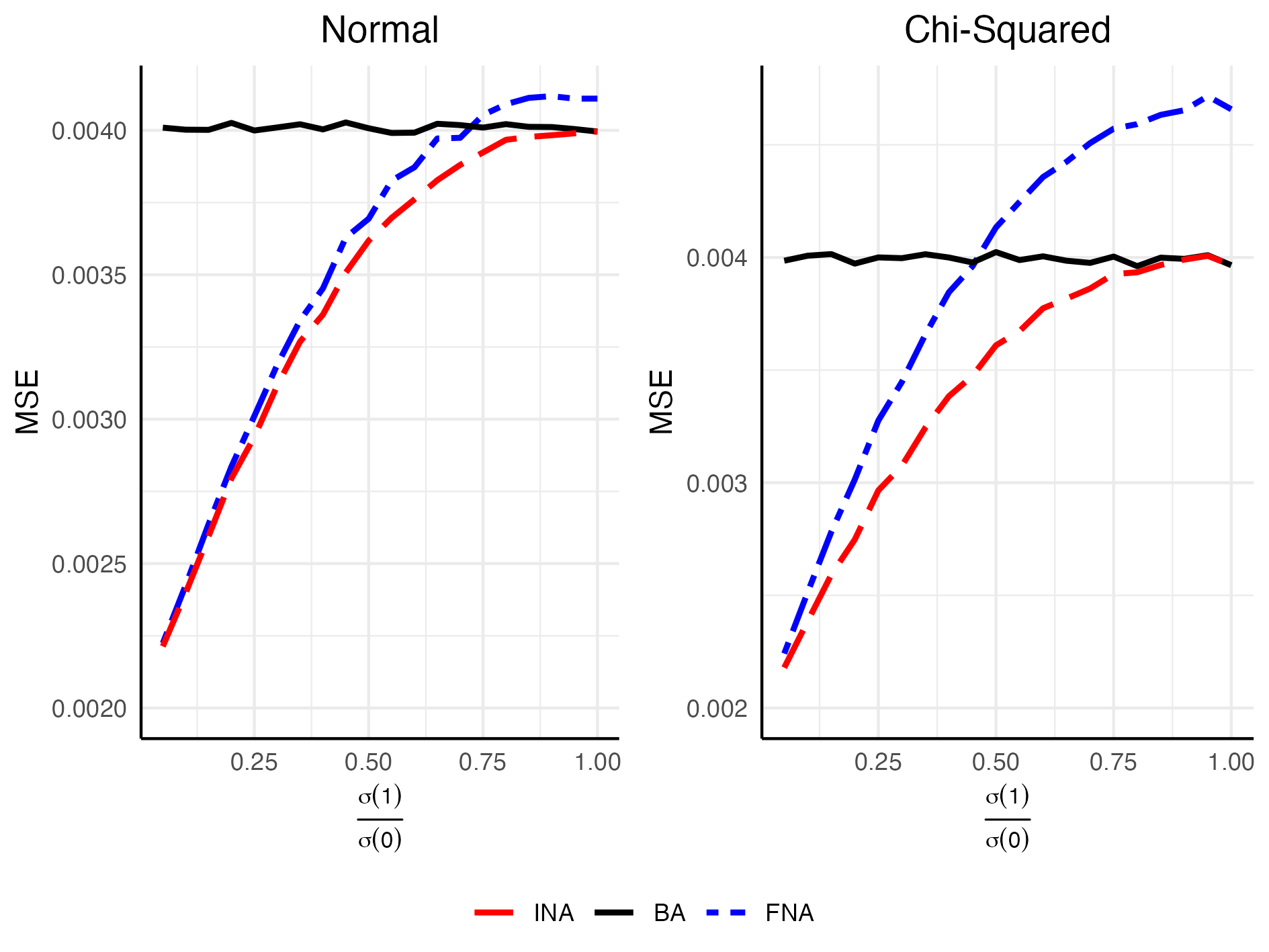}
\vspace{-3mm}
\caption{MSE under the Infeasible Neyman Allocation (INA), Balanced Allocation (BA) and Feasible Neyman Allocation (FNA) when $\varepsilon(1), \varepsilon(0)$ has the standard Normal distribution and the standardized Chi-Squared distribution. $m = 20$.}
\label{fig:mse12}
\end{figure}

\begin{figure}[htpb]
\centering
\includegraphics[width=1\linewidth]{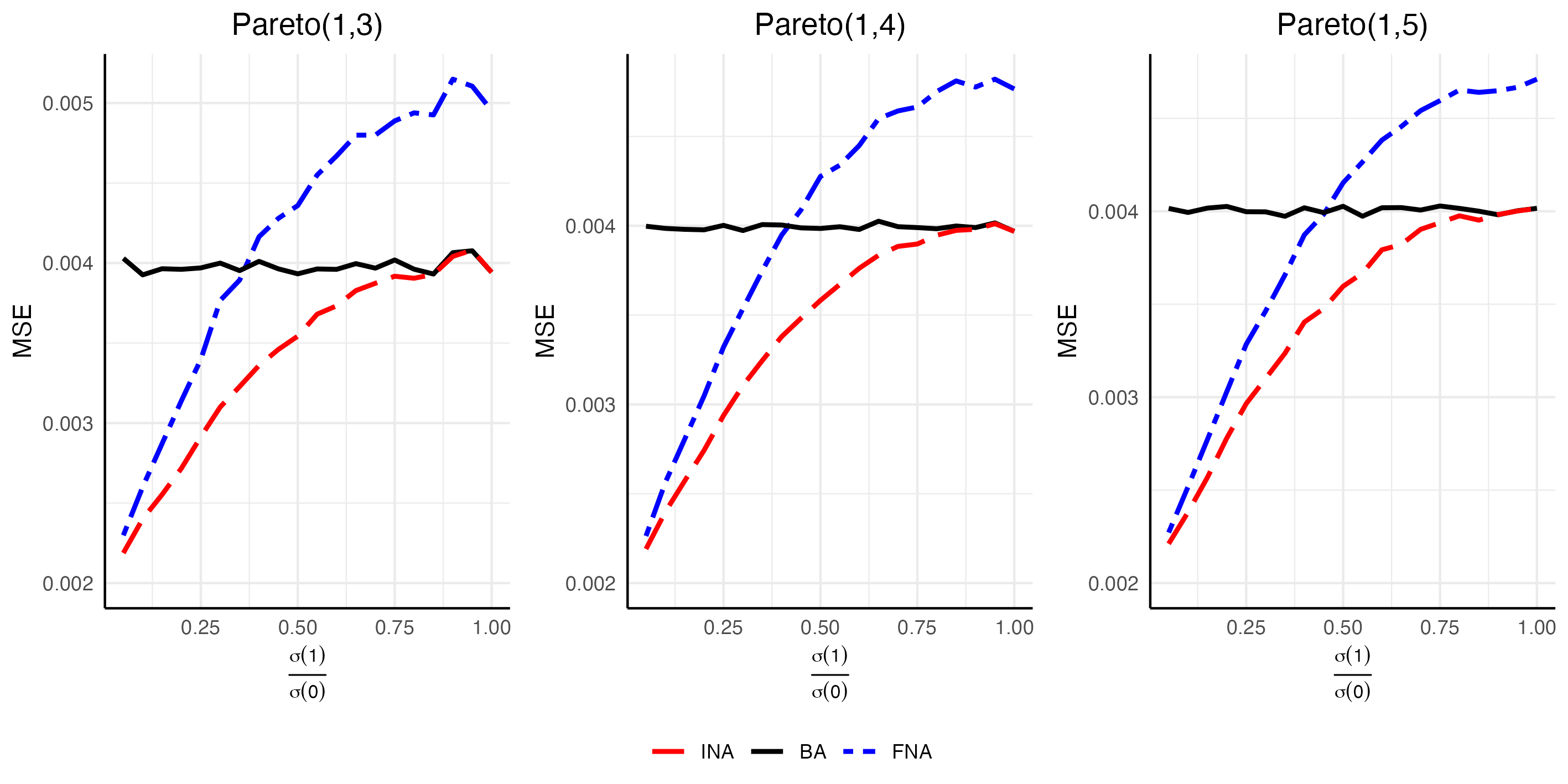}
\vspace{-8mm}
\caption{MSE under the Infeasible Neyman Allocation (INA), Balanced Allocation (BA) and Feasible Neyman Allocation (FNA) when $\varepsilon(1), \varepsilon(0)$ has the standardized Pareto($1,s$) distribution, with $s \in \{3,4,5\}$. $m = 20$.}
\label{fig:mse345}
\end{figure}

Results for Models 1 and 2 are presented in Figure \ref{fig:mse12}, while those for Models 3, 4 and 5 are presented in Figure \ref{fig:mse345}. First note that the MSE curves follow the theoretical description provided in Section \ref{section--implications}. The MSE of the FNA is monotonically increasing in $\sigma(1)/\sigma(0)$. As $\sigma(1)/\sigma(0) \to 0$, the MSE of $\hat{\theta}_{\tilde{p}}$ approaches 1/2 that of $\hat{\theta}_{\frac{1}{2}}$. As $\sigma(1)/\sigma(0) \to 1$, the MSE of $\hat{\theta}_{\tilde{p}}$ exceeds that of $\hat{\theta}_{\frac{1}{2}}$ by $B_m$.

Next, consider Model 1. With a normal distribution, the MSE of the FNA is 2.9\%
larger than the balanced allocation when $\sigma(1)/\sigma(0) = 1$. Weighed
against a potential reduction of 50\%, the FNA is definitely preferred in this
setting. However, the picture becomes less clear once we move to Model 2. Here,
the MSE of the FNA is around 17\% larger than that of the balanced
allocation. This is true up to $\sigma(1)/\sigma(0) = 0.75$, which is more
heteroskedasticity in many of the outcomes considered in our empirical
examples. Similar results obtain with Models 3, 4 and 5, in which the FNA has
MSEs that are 26\%, 20\% and 17\% more that of the balanced allocation when
$\sigma(1)/\sigma(0) = 1$. The above results suggest that the FNA may not be
advisable when $m = 20$. On the other hand, the worst-case efficiency loss
shrinks quickly as $m$ increases, so that the FNA performs favorably. Results
for $m = 50$ can be found in Appendix \ref{appendix--MSE}. At $m = 50$, the loss
in efficiency over Models 1 to 5 are, respectively, 1.5\%, 6.4\%, 13\%, 11\% and
8.6\%.

\section{Empirical Evidence of Approximate Homoskedasticity}
\label{section--empirical_homosked}

To assess the amount of heteroskedasticity that empirical researchers face, we
revisit the first 10 completed experiments in the AER RCT Registry. In each
experiment, we ask the following question: suppose the authors had access to a
small pilot prior to the main study, would they have done better using the FNA
instead of balanced randomization? In each experiment presented, we use the full
experimental sample to estimate the standard deviations of each treatment arm
and compute the corresponding ratios to see if these are close to one. In
practice, researchers cannot do this given a small pilot since they do not have
access to consistent variance estimators. Our findings suggest that these
authors would likely not have done better with the FNA. We present two examples
in this section: \citet{avvisati2014getting} conduct an experiment in which the
outcomes are relatively homoskedastic. \citet{ashraf2006tying} study outcome
variables which are heteroskedastic. In this latter example, we also provide
estimates of the interval \(C_{m}\) and show that it will be wide even when
\(m\) is large. The remaining eight experiments are qualitatively similar to
\citet{avvisati2014getting} and can be found in Section
\ref{section--empirical_additional} of the online appendix.

\subsection{\citet{avvisati2014getting}}

There is a significant body of research examining the impact of school-level
factors such as class size or teacher quality on educational performance of
students. These are typically seen as the primary instruments for educational
policy intervention. A large body of work also examines the impact of parental
inputs on educational outcomes. \citet{avvisati2014getting} study whether or
not parental inputs can be effectively manipulated through simple participation
programs at schools. They do so via a large-scale randomized control trial in
middle schools in the Cr\'eteil educational district of Paris. The experiment
targeted families of 6th graders and the program consisted of a sequence of
three meetings with parents every 2--3 weeks. The sessions focused on how
parents can help their children by participating at school and at home in their
education and additionally included advice on how to adapt to results in
end-of-term report cards. Participation in the program was randomized at the
class level -- half of the classes at a given school were assigned to the
participation program. Classes were groups of 20--30 students. The overall
sample comprised 183 classes and a total of 4,308 students. The study tracked
three types of outcomes: (1) parental involvement attitudes and behavior; (2)
children's behavior, namely truancy, disciplinary record and work effort; and
(3) children's academic results. Since randomization was done at the
class-level, we examine heteroskedasticity with respect to treatment status at
both the individual level and at the class level.

Table \ref{table:empirical--avvisati--student} reports student-level standard
deviations in treatment and control groups, as well as their ratios, for the
main outcomes of interest. These are the outcomes considered in Tables 2, 3 and
5 in \citet{avvisati2014getting}. The ratios are all close to one, so that by
and large the treatment and control groups are relatively homoskedastic at the
student level. This indicates that if the experimenters were to run a randomized
control trial in the same population with treatment assigned at the
individual-level, the FNA would likely yield no improvement relative to the
balanced allocation.

Table \ref{table:empirical--avvisati--class} reports standard deviations and
their ratios in class-level means of outcomes. This corresponds to a scenario in
which classes are the units of interest, with class-level means as the relevant
outcomes. We first calculate class-level means and then compute their standard
deviations across classes for the treatment and control groups respectively. The
standard deviation ratios for class-level means are by and large also close to
one. Hence, if the hypothetical experiment were to be conducted at the
class-level, the treatment and control groups would still be relatively
homoskedastic. In this case, the FNA would again not improve upon the balanced
allocation.

\begin{table}[htbp]
  \centering\small
  \caption{Student-Level Heteroskedasticity in \citet{avvisati2014getting}.}
    \begin{tabular}{clccc}\hline\hline
    \multicolumn{2}{c}{Outcome Variables} & Treatment & Control & Treat./Cont.\\
    \midrule
    \multirow{6}[0]{*}{ \shortstack{Parental \\ Involvement}}
          & Global parenting score                & 0.34  & 0.34  & 1.01 \\
          & School-based involvement score        & 0.66  & 0.63  & 1.05 \\
          & Home-based involvement score          & 0.59  & 0.57  & 1.04 \\
          & Understanding and perceptions score   & 0.53  & 0.55  & 0.97 \\
          & Parent-school interaction             & 0.40  & 0.40  & 1.01 \\
          & Parental monitoring of school work    & 0.43  & 0.41  & 1.05 \\
    \midrule
    \multirow{7}[0]{*}{Behavior} & Absenteeism    & 6.29  & 8.63  & 0.73 \\
          & Pedagogical team: Behavioral score    & 0.73  & 0.74  & 0.98 \\
          & Pedagogical team: Discipl. sanctions  & 1.20  & 1.18  & 1.02 \\
          & Pedagogical team: Good conduct        & 0.49  & 0.47  & 1.04 \\
          & Pedagogical team: Honors              & 0.28  & 0.32  & 0.89 \\
          & Teacher assessment: Behavior in class & 0.48  & 0.49  & 0.98 \\
          & Teacher assessment: School work       & 0.49  & 0.50  & 1.00 \\
    \midrule
    \multirow{6}[0]{*}{Test Scores} & French (Class grade)
                                                  & 3.73  & 3.70  & 1.01 \\
          & Mathematics (Class grade)             & 4.26  & 4.25  & 1.00 \\
          & Average across subjects (Class grade) & 2.87  & 2.88  & 1.00 \\
          & Progress over the school year         & 0.49  & 0.49  & 0.99 \\
          & French (Uniform test)                 & 0.99  & 1.01  & 0.98 \\
          & Mathematics (Uniform test)            & 0.99  & 1.02  & 0.98 \\
    \hline\hline
    \end{tabular}%
  \label{table:empirical--avvisati--student}%
\end{table}%

\begin{table}[htbp]
  \centering\small
  \caption{Class-Level Heteroskedasticity in \citet{avvisati2014getting}.}
    \begin{tabular}{clccc}\hline\hline
    \multicolumn{2}{c}{Outcome Variables} & Treatment & Control & Treat./Cont.\\
    \midrule
        \multirow{6}[0]{*}{ \shortstack{Parental \\ Involvement}}
              & Global parenting score                & 0.12  & 0.12  & 0.97 \\
              & School-based involvement score        & 0.24  & 0.35  & 0.69 \\
              & Home-based involvement score          & 0.20  & 0.15  & 1.33 \\
              & Understanding and perceptions score   & 0.20  & 0.22  & 0.94 \\
              & Parent-school interaction             & 0.13  & 0.12  & 1.09 \\
              & Parental monitoring of school work    & 0.13  & 0.14  & 0.96 \\
        \midrule
        \multirow{7}[0]{*}{Behavior} & Absenteeism    & 2.21  & 3.45  & 0.64 \\
              & Pedagogical team: Behavioral score    & 0.24  & 0.27  & 0.88 \\
              & Pedagogical team: Discipl. sanctions  & 0.36  & 0.36  & 1.01 \\
              & Pedagogical team: Good conduct        & 0.22  & 0.23  & 0.95 \\
              & Pedagogical team: Honors              & 0.10  & 0.11  & 0.87 \\
              & Teacher assessment: Behavior in class & 0.16  & 0.16  & 1.01 \\
              & Teacher assessment: School work       & 0.15  & 0.14  & 1.02 \\
        \midrule
        \multirow{6}[0]{*}{Test Scores}
              & French (Class grade)                  & 1.32  & 1.31  & 1.01 \\
              & Mathematics (Class grade)             & 1.76  & 1.84  & 0.96 \\
              & Average across subjects (Class grade) & 0.83  & 0.93  & 0.89 \\
              & Progress over the school year         & 0.16  & 0.14  & 1.11 \\
              & French (Uniform test)                 & 0.42  & 0.42  & 1.01 \\
              & Mathematics (Uniform test)            & 0.40  & 0.43  & 0.93 \\
              \hline\hline
    \end{tabular}%
  \label{table:empirical--avvisati--class}%
\end{table}%

\subsection{\citet{ashraf2006tying}}

A large body of economic models posit that individuals have time inconsistent
preferences, exhibiting more impatience for near-term trade-offs than for future
trade-offs. The implication of these models is that those who engage in
commitment devices ex ante may improve their welfare. To test this hypothesis,
\citet{ashraf2006tying} conducted an RCT in the Philippines, in which
individuals were offered randomly offered the chance to open a SEED (Save, Earn,
Enjoy Deposits) account. Money deposited into the account cannot be withdrawn
until the owner reached a goal, such as reaching a savings amount or until a
pre-specified month in which they expected large expenditures.

Partnering with a rural bank in Mindanao, the authors surveyed 1,777 of their
existing or former clients, of which 842 were placed into the treatment group,
while 469 were placed in the control group.  As treatment involved receiving a
briefing on the importance of savings, the remaining 466 individuals were placed
in the marketing group, receiving the briefing but not access to SEED. We focus
on the Table VI of \citet{ashraf2006tying}, containing results on saving
behavior. The parameter of interest is the Intent-to-Treat effect, with
approximately 25\% of the treated taking up treatment. Here, the authors find
that relative to the control group, the treatment group had a higher change in
savings 6 months (6m) and 12 months (12m) after treatment. Comparing the
treatment to the marketing group led to weaker but still positive results.

We present standard deviations of the outcomes as well as their ratios in Table
\ref{table:empirical--ashraf}. The outcomes of interest are
\begin{enumerate}
\item Change in Total Balance (6m) (\emph{\(\Delta\) Tot. Bal. (6m)}),
\item Change in Total Balance (12m) (\emph{\(\Delta\) Tot. Bal. (12m)}),
\item Change in Total Balance exceeds 0\% (12m) (\emph{\(\Delta\) Tot. Bal. $>
  0\%$ (12m)}),
\item Change in Total Balance exceeds 20\% (12m) (\emph{\(\Delta\) Tot. Bal. $>
  20\%$ (12m)}).
\end{enumerate}
We first note that \emph{\(\Delta\) Tot. Bal. $> 0\%$ (12m)} and
\emph{\(\Delta\) Tot. Bal. $> 20\%$ (12m)} are binary outcomes which are
relatively homoskedastic. \emph{\(\Delta\) Tot. Bal. (6m)} and \emph{\(\Delta\)
Tot. Bal. (12m)}, measured in Philippine pesos, exhibit more
heteroskedasticity. In particular, comparing the treatment group to the
marketing group at the 12 month period, we observe a standard deviation ratio of
$3.13$. At first glance, this suggests that the FNA might outperform balanced
randomization, at least with respect to this specific outcome. This turns out to
be false once we investigate other features of \emph{\(\Delta\)
Tot. Bal. (6m)} and \emph{\(\Delta\) Tot. Bal. (12m)}. Quantiles of these
variables are displayed in Table \ref{table:empirical--ashraf--quantiles}.
Clearly, they have extremely fat right
tails, which we confirm by computing the kurtosis, contained in Table
\ref{table:empirical--ashraf--kurtosis}. Fat tails worsen the performance of a
variety of statistical techniques, including the FNA, as our analysis in Section
\ref{section--theoretical-results} shows.

\begin{table}[htbp]
  \centering
  \caption{Heteroskedasticity in \citet{ashraf2006tying}}
  \begin{tabular}{lccccc}\hline\hline
    & Treat. & Cont. & Market. & Treat./Cont. & Treat./Market. \\
    \midrule
    \(\Delta\) Tot. Bal. (6m) & 2347.60 & 2880.70 & 1335.98 & 1.76  & 0.81 \\
    \(\Delta\) Tot. Bal. (12m) & 6093.24 & 1945.00 & 2690.65 & 2.26  & 3.13 \\
    \(\Delta\) Tot. Bal. $>$ 0\% (12m) & 0.47  & 0.45  & 0.42  & 1.11  & 1.05 \\
    \(\Delta\) Tot. Bal. $>$ 20\% (12m) & 0.40  & 0.35  & 0.31  & 1.30  & 1.16 \\
    \hline\hline
  \end{tabular}%
  \label{table:empirical--ashraf}%
\end{table}%

\begin{table}[htbp]
  \centering
  \caption{Quantiles of Outcome Variables in \citet{ashraf2006tying}.}
  \begin{tabular}{ccccccccccc}\hline\hline
    Variable & Group & 1\%   & 5\%   & 10\%  & 50\%  & 90\%  & 95\%  & 99\%  & 99.5\% & 99.9\% \\
    \midrule
    \multicolumn{1}{c}{\multirow{3}[-1]{*}{\shortstack{\(\Delta\)\\Tot. Bal.\\(6m)}}} & Treat. & -1100 & -500  & -300  & 0     & 500   & 1500  & 7200  & 13100 & 28900 \\
    & Market. & -1000 & -600  & -400  & 0     & 100   & 900   & 5600  & 12200 & 40800 \\
    & Cont. & -1600 & -600  & -500  & 0     & 0     & 600   & 2700  & 6100  & 18400 \\
    \midrule
    \multicolumn{1}{c}{\multirow{3}[-1]{*}{\shortstack{\(\Delta\)\\Tot. Bal.\\(12m)}}} & Treat. & -1300 & -900  & -500  & 0     & 500   & 1600  & 8500  & 18100 & 102300 \\
    & Market. & -1300 & -900  & -500  & -100  & 300   & 1600  & 10500 & 15700 & 19900 \\
    & Cont. & -2000 & -1200 & -800  & -100  & 100   & 900   & 6500  & 8100  & 34300 \\
    \hline\hline
  \end{tabular}%
  \label{table:empirical--ashraf--quantiles}%
\end{table}%

\begin{table}[htbp]
  \centering
  \caption{Kurtosis of the Outcome Variables in \citet{ashraf2006tying}}
  \begin{tabular}{lccc}\hline\hline
    Variable & Treat. & Market. & Cont. \\
    \midrule
    \(\Delta\) Tot. Bal. (6m) & 252.78 & 218.92 & 156.33 \\
    \(\Delta\) Tot. Bal. (12m) & 258.56 & 66.56 & 309.89 \\
    \hline\hline
  \end{tabular}%
  \label{table:empirical--ashraf--kurtosis}%
\end{table}%

\begin{figure}[htbp]
  \centering
  \includegraphics[width=1\linewidth]{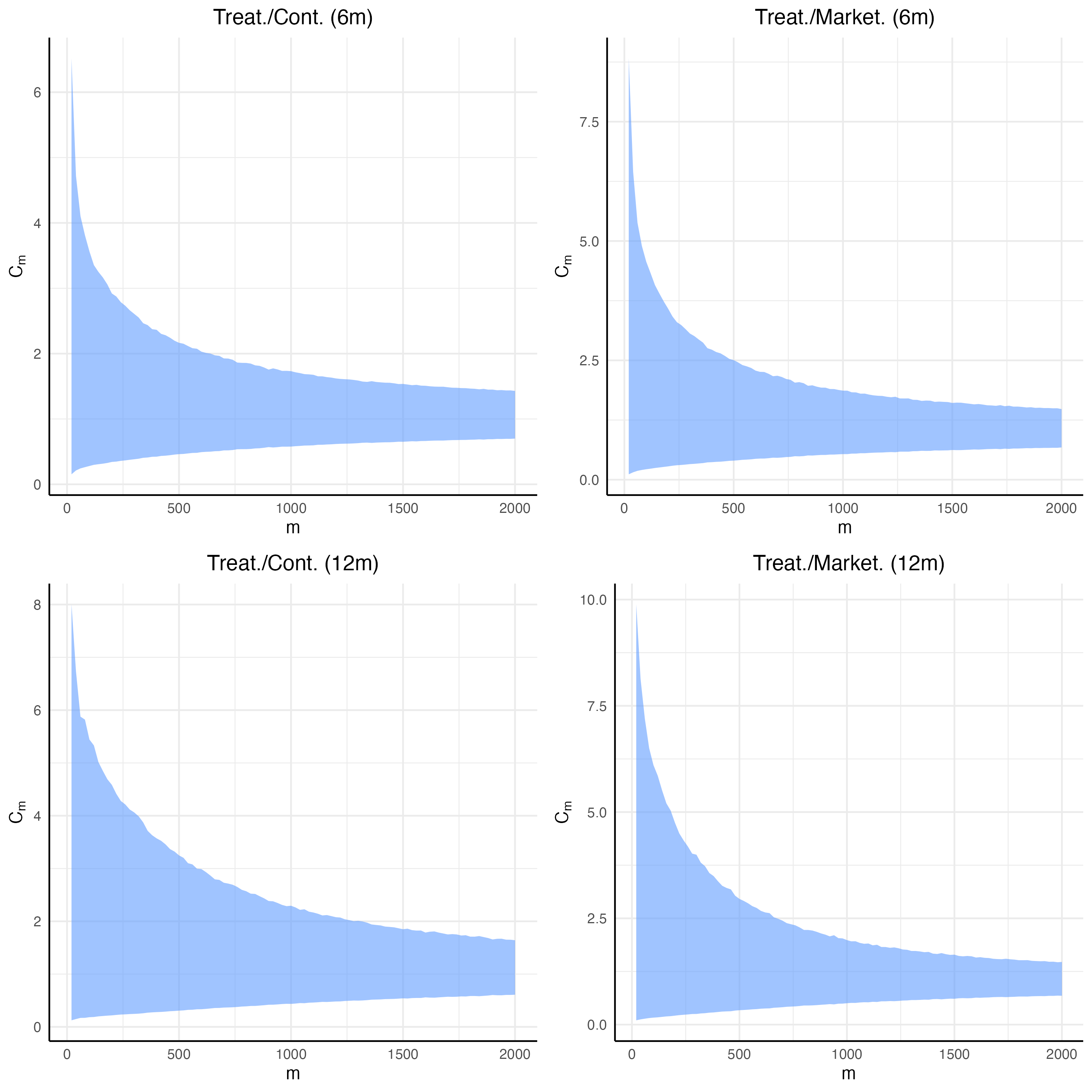}
  \caption{$C_m$ for the outcomes of interest estimated using the full
  experiment data. Interval computed under the assumption that the pilot assigns
  half of the units to treatment and half to control.}
  \label{fig:empirical--ashraf--cm}
\end{figure}

Researchers cannot estimate $C_m$ if they only have access to a small
pilot. However, this is feasible with data from the full experiment. To do so, we draw samples of size $m$ with replacement and compute $Z_m$ for each draw. We then take the average of $Z_m + 1/Z_m$ over 10,000 draws to evaluate $B_m$, essentially evaluating the expectation with respect to the empirical CDF. Proposition \ref{theorem--Cm_small_m} then allows us to compute $C_m$. We evaluate $C_m$ for $m \leq 2000$, using grids of size $20$. Results are displayed in Figure
\ref{fig:empirical--ashraf--cm}. Given the high kurtosis, there is a relatively
large range of ratios of standard deviations for which the balanced allocation
is preferred to the FNA.

We can further compute the $m$ necessary before the FNA
outperforms the balanced allocation. Given the upper and lower bounds in Figure \ref{fig:empirical--ashraf--cm}, we smooth over the grid points using cubic interpolation. We then find the minimum $m$ at which the either bound attains the values in Table \ref{table:empirical--ashraf}. These are the ``Exact" intervals presented in Table \ref{table:empirical--ashraf--min_size}.
Here, we see that the necessary pilot sizes are between 25--50\% of the full experiment. For
the comparison of the treatment and marketing groups at 6 months, the necessary pilot
size exceeds $2,000$, falling outside the set of grid points we explored. To
complete the analysis, we use Proposition \ref{theorem--Cm_large_m} to obtain approximations of
the necessary $m$. Specifically, we estimate $V$ using their sample analogs computed on the full experiment:
\begin{equation*}
  \hat{V} = \frac{\frac{1}{n_1}\sum_{A_i = 1} \left(Y_i - \overline{Y}(1)\right)^4
  }{\left(\frac{1}{n_1-1} \sum_{A_i = 1} \left(Y_i - \overline{Y}(1)\right)^4
  \right)^2} + \frac{\frac{1}{n_0}\sum_{A_i = 0} \left(Y_i -
  \overline{Y}(0)\right)^4 }{\left(\frac{1}{n_0-1} \sum_{A_i = 0} \left(Y_i -
  \overline{Y}(0)\right)^4 \right)^2} - 2 ~,
\end{equation*}
where $n_a$ is the number of units with treatment status $a$ in the full
experiment and \(\overline{Y}(a) = \frac{1}{n_a} \sum_{A_i = a} Y_i\). We can then estimate $m$ by
\begin{equation*}
  \hat{m} = \frac{\hat{V}}{\left(1 - \frac{\hat{\sigma}(1)}{\hat{\sigma}(0)}\right)^2}~.
\end{equation*}

Comparing the asymptotic intervals to the exact ones, we see
that the former is far too optimistic for the fat-tailed DGP in
\citet{ashraf2006tying}. This is likely because the sub-Gaussian assumption is
inappropriate for this data. Nonetheless, applying the asymptotic bounds to the
case of the treatment vs marketing groups at 6 months, we find that a pilot size of
7,000 would be necessary for the FNA to outperform balanced randomization. This
is nearly 4 times the size of the actual experiment.

\begin{table}[htbp]
  \centering
  \caption{Necessary Pilot Sizes}
    \begin{tabular}{lcc}\hline\hline
          & \multicolumn{1}{l}{Treat./Cont.} & \multicolumn{1}{l}{Treat./Market.} \\
    \midrule
    Exact 6m & 930.70 & -- \\
    Exact 12m & 1014.94 & 475.29 \\
    \midrule
    Asympt. 6m & 355.02 & 6857.59 \\
    Asympt. 12m & 177.10 & 35.52 \\
    \hline\hline
    \end{tabular}%
  \label{table:empirical--ashraf--min_size}%
\end{table}%

In sum, our analysis suggests that the FNA would have performed poorly in the
context of \citet{ashraf2006tying}. Even though the outcomes of interest exhibit
stronger heteroskedasticity, the fat tails of the outcome distributions also
impedes the estimation of $\widetilde{p}$, so that ultimately, very large
pilot sizes are needed for the FNA to outperform the balanced allocation in this
example.

\section{Potential Solutions}

In this section, we use simulations to explore methods that can improve the
performance of the FNA with small pilots. For definitions of the models, see
Section \ref{section--mse_sims}. We consider in turn (1) testing for
homoskedasticity, (2) additive regularization and (3) exponential
regularization. We conclude with a comparison of these results and speculate
that exponential regularization is preferable to the other two
methods. Additional simulation results can be found in Appendix
\ref{appendix--sim_solutions}.

\subsection{Testing for Homoskedasticity}

Since balanced randomization is preferred when outcomes are relatively homoskedastic, consider the following two-step procedure. First test for homoskedasticity in the pilot. If the test rejects, use the FNA in the main wave. Otherwise, use the balanced allocation instead.

Specifically, we will use the (linear) Wald test for the equality of variance, which has the following test statistic:
\begin{equation*}
  \tilde{W} = \sqrt{\frac{m}{2}}\frac{(\tilde{\sigma}^2(1) -
  \tilde{\sigma}^2(0))}{\sqrt{\tilde{V}_W(1)+ \tilde{V}_W(0)}}~,
\end{equation*}
where
\begin{gather*}
  \tilde{V}_W(a) = \begin{pmatrix}
    1 & - 2 \overline{\widetilde{Y}}(a)
  \end{pmatrix} \widetilde{\Sigma}_{W} (a) \begin{pmatrix}
    1 \\
    -2 \overline{\widetilde{Y}}(a)
  \end{pmatrix} ~, \\
  \widetilde{\Sigma}_{W} (a) = \begin{pmatrix}
    \frac{1}{m/2} \sum_{\widetilde{A}_i = a} \left( \widetilde{Y}_i^2 -
    \overline{\widetilde{Y}^2} (a) \right)^2 &
    \frac{1}{m/2} \sum_{\widetilde{A}_i = a} \left( \widetilde{Y}_i -
    \overline{\widetilde{Y}} (a) \right) \left( \widetilde{Y}_i^2 -
    \overline{\widetilde{Y}^2} (a) \right) \\
    * & \frac{1}{m/2} \sum_{A_i = a} (\widetilde{Y}_i -
    \overline{\widetilde{Y}} (a))^2
  \end{pmatrix} ~, \\
  \overline{\widetilde{Y}}(a) = \frac{1}{m/2}\sum_{A_i = a} \widetilde{Y}_i
  \quad , \quad \overline{\widetilde{{Y}}^2}(a) = \frac{1}{m/2}\sum_{A_i = a}
  \widetilde{Y}_i^2 ~.
\end{gather*}
It is well known that $\tilde{W} \overset{d}{\to} N(0,1)$ as $m \to \infty$.
We can therefore form a two-sided, level $\alpha$ test for homoskedasticity with the rejection rule:
\begin{equation*}
  \phi = \mathbf{1}\left\{\tilde{W} > \Phi^{-1}\left(1-\frac{\alpha}{2}\right) \right\} + \mathbf{1}\left\{\tilde{W} < \Phi^{-1}\left(\frac{\alpha}{2}\right) \right\}~,
\end{equation*}
where $\Phi(\cdot)$ is the CDF of the standard normal distribution. Under small pilot asymptotics, this test will not be consistent. However, the two-step procedure will be asymptotically equivalent to the FNA when pilots are large, since for any fixed level $\alpha$, the test will reject with probability approaching $1$ as $m \to \infty$. However, under such a ``pre-testing procedure", the main-wave difference-in-means estimator will not be uniformly consistent.

\begin{figure}[htpb]
\centering
\vspace{-5mm}
\includegraphics[width=0.8\linewidth]{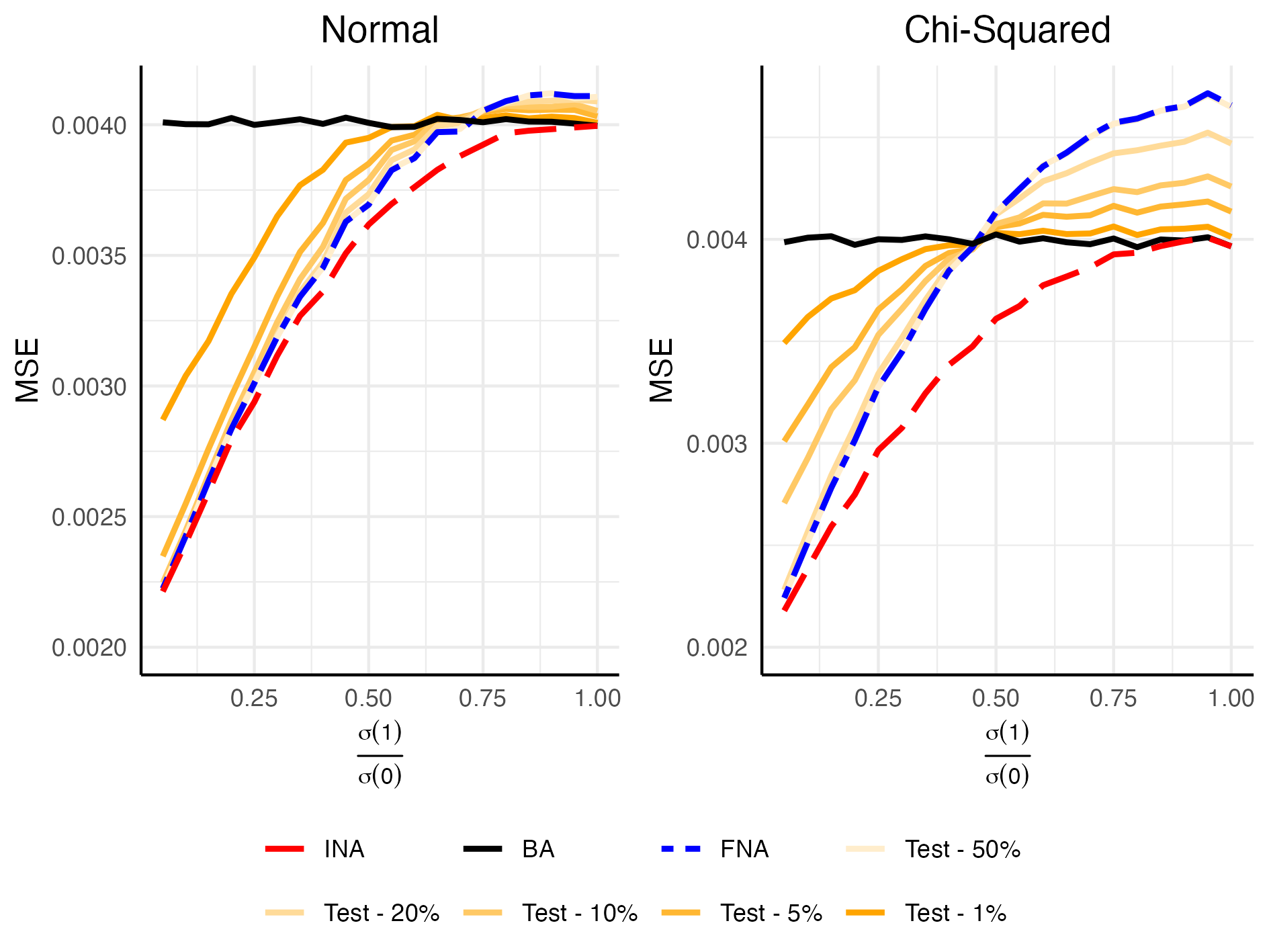}
\vspace{-3mm}
\caption{MSE under testing for homoskedasticity. We compare the Infeasible Neyman Allocation (INA), Balanced Allocation (BA), Feasible Neyman Allocation (FNA), as well as testing for homoskedasticity with various values of $\alpha$ (Test - $\alpha\%$).}
\label{fig:mse12_sol_test}
\vspace{5mm}
\includegraphics[width=0.8\linewidth]{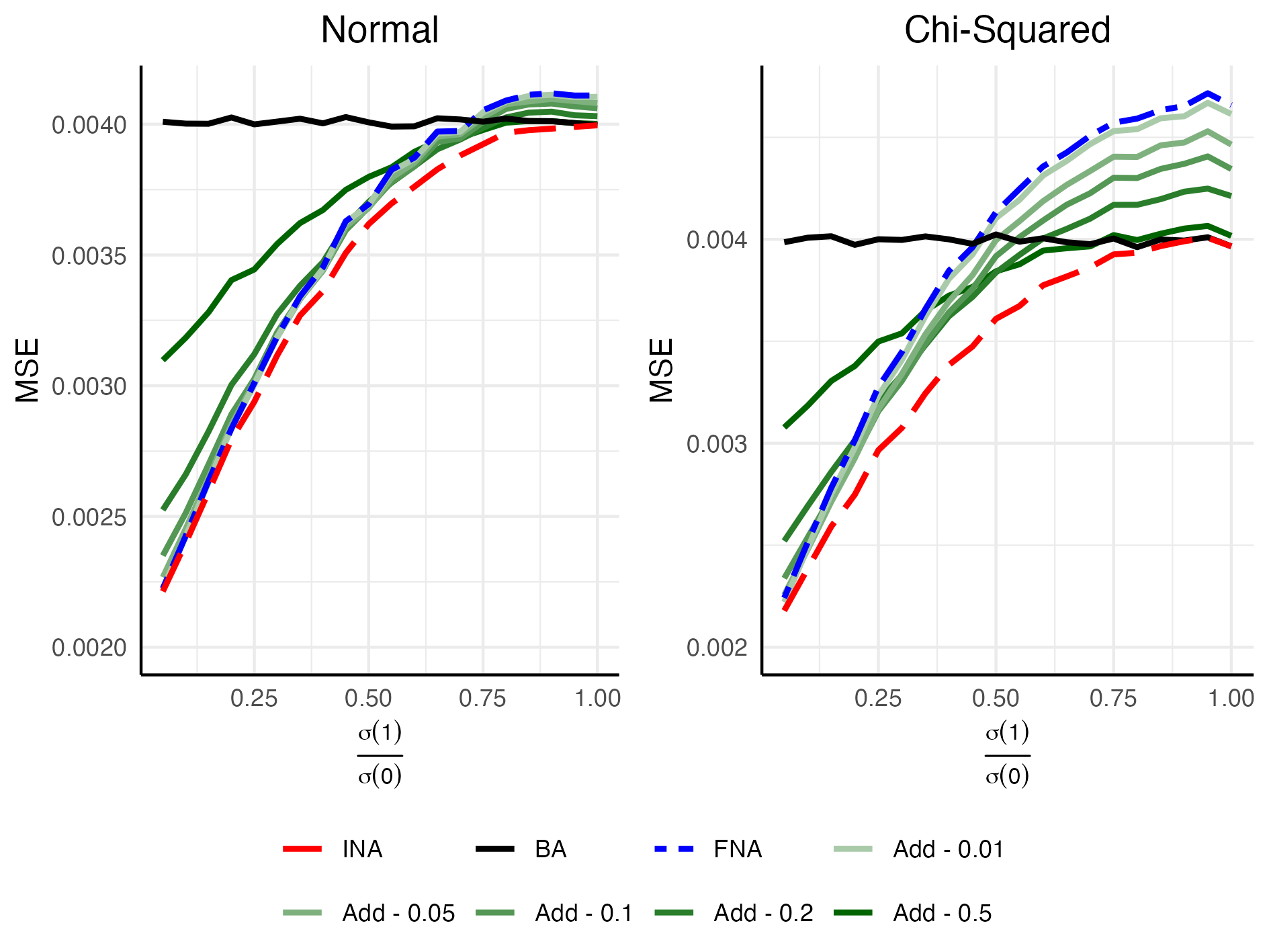}
\vspace{-3mm}
\caption{MSE under additive regularization. We compare the Infeasible Neyman Allocation (INA), Balanced Allocation (BA), Feasible Neyman Allocation (FNA), as well as additive regularization with various values of $\nu$ (Add - $\nu$).}
\label{fig:mse12_sol_add}
\end{figure}

{\sloppy
Nonetheless, the procedure appears to perform well in practice. Figure \ref{fig:mse12_sol_test} presents the MSEs that obtain using first stage tests with $\alpha \in \left\{1\%, 5\%, 10\%, 20\%, 50\%\right\}$. When $\sigma(1)/\sigma(0) \geq 0.5$, decreasing $\alpha$ decreases MSE across all distributions. Conversely, when $\sigma(1)/\sigma(0) \leq 0.5$, decreasing $\alpha$ increases MSE across all distributions. The appropriate level of $\alpha$ then depends on researcher's preference for the performance of the method over the range of $\sigma(1)/\sigma(0)$.}

\subsection{Additive Regularization}

Rather than using a test to make a binary decision between the FNA and the balanced allocation, we can consider regularizing the FNA towards the balanced allocation. In particular, for a given tuning parameter $\nu$, let
\begin{equation*}
  \tilde{p}^\text{add} = \frac{\tilde{\sigma}(1) + \tilde{\sigma}(0)\nu}{(\tilde{\sigma}(1) + \tilde{\sigma}(0))(1+\nu)}~.
\end{equation*}
In other words, we regularize $\tilde{\sigma}(1)$ by adding to it $\nu \tilde{\sigma}(0)$, and similarly for $\tilde{\sigma}(0)$.
Pre-multiplying $\nu$ with $\tilde{\sigma}(1)$ and $\tilde{\sigma}(1)$ ensures that $\tilde{p}^\text{add}$ is invariant to their scale. Furthermore,
\begin{align*}
  \tilde{p}^\text{add} \to 1/(1+\nu) &\mbox{ as }  \sigma(1)/\sigma(0) \to \infty~, \\
    \tilde{p}^\text{add} \to \nu/(1+\nu) &\mbox{ as }  \sigma(1)/\sigma(0) \to 0~.
\end{align*}
Figure \ref{fig:mse12_sol_add} presents results for $\nu \in \{0.01, 0.05, 0.1,
0.2, 0.5\}$. As we increase $\nu$, the MSE decreases for the homoskedastic
region but increases for the heteroskedastic region.

\subsection{Exponential Regularization}

Another method by which we can continuously regularize the FNA is as follows. For a given tuning parameter $\tau$, let
\begin{equation*}
  \tilde{p}^{\text{exp}} = \frac{\left(\frac{\tilde{\sigma}(1)}{\tilde{\sigma}(0)}\right)^\tau}{1+ \left(\frac{\tilde{\sigma}(1)}{\tilde{\sigma}(0)}\right)^\tau}~.
\end{equation*}
Observe that when $\tau = 1$, $\tilde{p}^{\text{add}}$ is FNA. For $0 \leq \tau
< 1$, $\tilde{p}^{\text{exp}}$ regularizes the FNA towards the balanced
allocation. In particular, for $\tau = 0$, $\tilde{p}^{\text{add}} = \frac{1}{2}$.
Figure \ref{fig:mse12_sol_add} presents results for $\tau \in \{0.99, 0.95, 0.9,
0.8, 0.5\}$. As before, when we decrease $\tau$, the MSE decreases for the homoskedastic region but increases for the heteroskedastic region.

\begin{figure}[htpb]
\vspace{-5mm}
\centering
\includegraphics[width=0.8\linewidth]{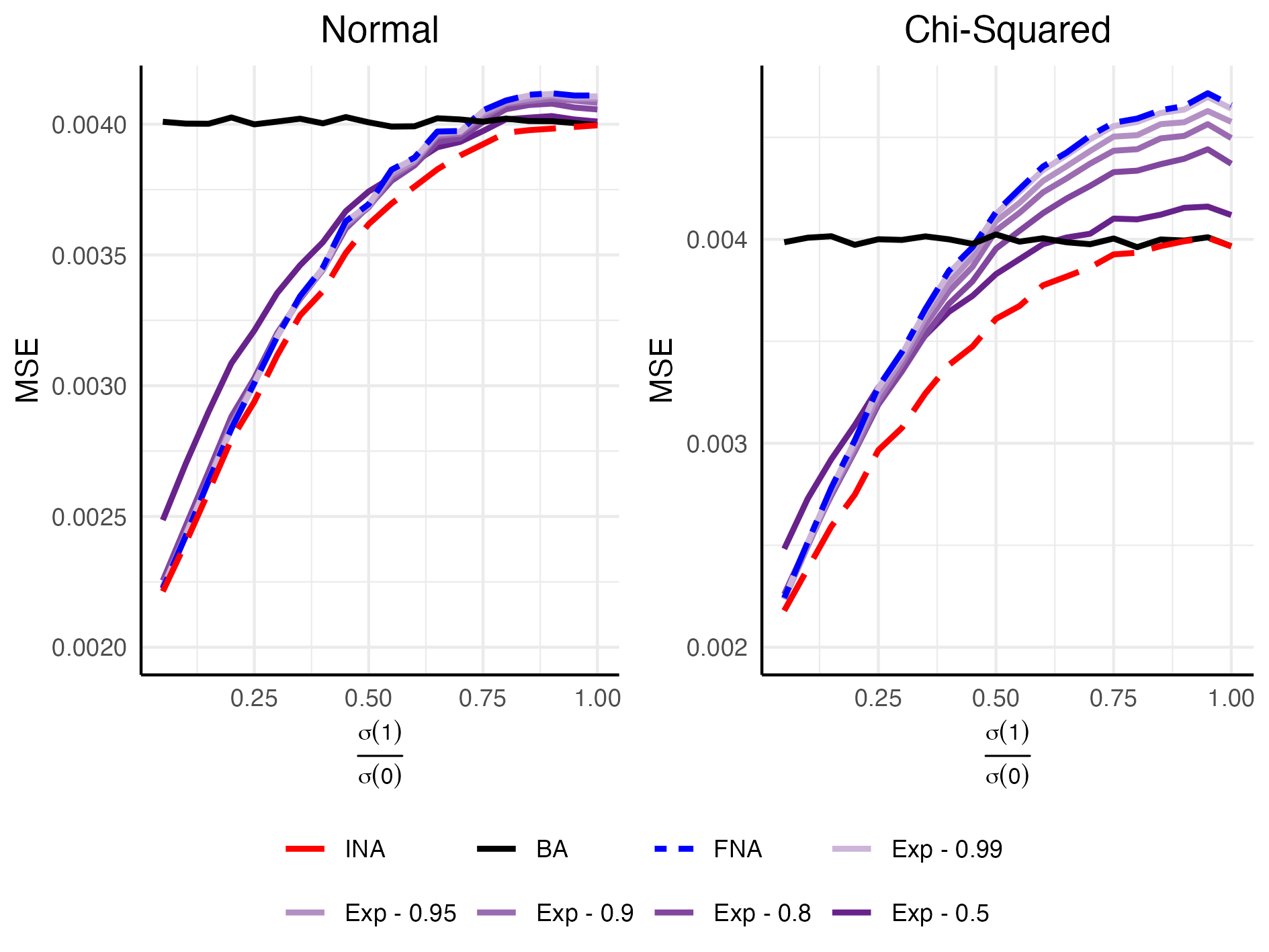}
\vspace{-3mm}
\caption{MSE under exponential regularization. We compare the Infeasible Neyman Allocation (INA), Balanced Allocation (BA), Feasible Neyman Allocation (FNA), as well as exponential regularization with various values of $\tau$ (Exp - $\tau$).}
\vspace{5mm}
\includegraphics[width=0.8\linewidth]{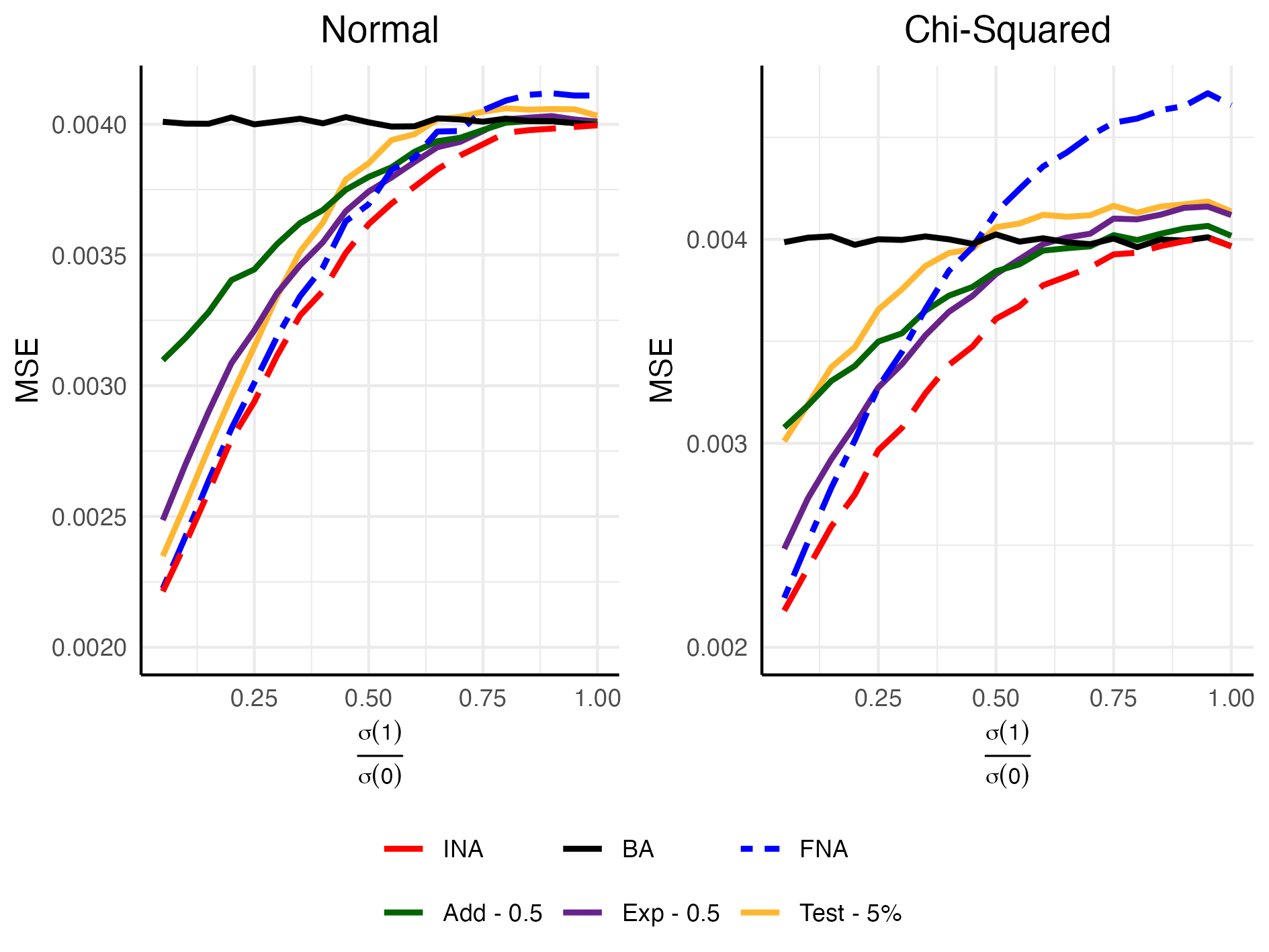}
\vspace{-3mm}
\caption{MSE under various candidate solutions. The tuning parameter for each solution is chosen so that their performance in the homoskedastic region is approximately comparable.}
\label{fig:mse12_sol_compare}
\end{figure}

\subsection{Comparison of Solutions}

As the above sections demonstrate, each potential solution entails different
trade-offs between MSE in the homoskedastic and heteroskedastic region.
These trade-offs also vary depending on the tuning parameter chosen. In a crude attempt at comparing these methods, we consider the following. Choose tuning parameter values so that the MSEs of the methods are made as close as possible in the homoskedastic region. Then compare MSEs in the heteroskedastic region.

One such comparison is displayed in Figure \ref{fig:mse12_sol_compare}. Here, we
see that having fixed performance in the homoskedastic region, exponential
regularization seems to do a better job in the heteroskedastic region. This
conclusion ostensibly extends to the Pareto distributions, as well as other choices of tuning parameters that keep the solutions appropriately comparable. We therefore speculate that exponential regularization is an effective method for improving the small sample performance of the FNA.

\section{Conclusion}
\label{section--conclusion}

We study the properties of the Feasible Neyman Allocation (FNA) in an asymptotic
framework for two-wave experiments that takes pilot size to be fixed as the size
of the main wave tends to infinity. In this setting, the estimated allocation
has error that is not negligible even in the limit. Our asymptotic model
therefore corresponds more closely to the finite sample statistical problem in
which pilots are small and the optimal allocation may be poorly estimated. We
characterize the conditions under which the difference-in-means estimator has
larger asymptotic variance compared to balanced randomization, where half of the
main wave is assigned to treatment and the remainder to control. This happens
when the potential outcomes are relatively homoskedastic with respect to
treatment status or exhibit high kurtosis -- situations that may arise in
practice. Our results suggest that the confluence of these factors -- small
pilots, homoskedasticity, heavy-tails -- are disadvantageous to the
FNA. Instead, researchers might find it useful to consider alternative
procedures such as regularizing the FNA towards the balanced allocation.

\section*{Acknowledgements}

We thank the editor, the associate editor and three anonymous referees for
comments and suggestions that greatly improved the article. We are also grateful
to Federico Bugni, Ivan Canay, Joel Horowitz, Dean Karlan, Max Tabord-Meehan and
Chris Udry for helpful discussions. We report that there are no competing
interests to declare.

\bibliographystyle{chicago}

\bibliography{2024_neyman_vs_kiss}

\newpage

\appendix

\setcounter{page}{1}

\begin{center}
  \Large \textbf{Online Appendix for ``On the Performance of the Neyman
  Allocation with Small Pilots''}
\end{center}

\section{Proofs}

\subsection{Proof of Corollary \ref{corollary--suboptimal_variance}}

Let \(\mathcal{L}_{m}\) denote a random variable whose distribution is given by
\[
\mathbb{P} \left( \mathcal{L}_{m} \leq t \right) = \int_{0}^{1} \Phi \left(
\frac{t}{s \left( p \right)} \right) \; G_{m} (\mathrm{d} p)~.
\]
Using the Law of Iterated Expectations, it can be shown that \(\mathcal{L}_{m}\) has mean zero and variance given by
\[
\mathbb{E} \left[ \mathcal{L}_{m}^{2} \right] = \mathbb{E} \left[ s^{2} \left(
\widetilde{p} \right) \right] = \mathbb{E} \left[\frac{\sigma^{2}
  (1)}{\widetilde{p}} + \frac{\sigma^{2} (0)}{1 - \widetilde{p}} \right]~.
\]
Next, note that for all $\widetilde{p} \in [0,1], \widetilde{p} \neq p_*$,
\begin{equation*}
  \frac{\sigma^{2}(1)}{\widetilde{p}} + \frac{\sigma^{2} (0)}{1 - \widetilde{p}}  >  \frac{\sigma^{2}(1)}{p_*} + \frac{\sigma^{2} (0)}{1 - p_*} = \Sigma_*~.
\end{equation*}
Since $\sigma^2(0) > 0$ and $\sigma^2(1) > 0$, $\mathbb{P}\left(\widetilde{p} \neq p_* \right) > 0$. That is,
\begin{equation*}
  \mathbb{P}\left( \frac{\sigma^{2}(1)}{\widetilde{p}} + \frac{\sigma^{2} (0)}{1 - \widetilde{p}}  > \Sigma_* \right) > 0~.
\end{equation*}
By the strict monotonicity of expectation, we are done. \qed

\subsection{Proof of Proposition \ref{theorem--Cm_small_m}}

By our earlier derivation, $\widehat{\theta}_p$ has lower asymptotic variance than $\widehat{\theta}_{\widetilde{p}}$ if and only if
\begin{align*}
  \mathbb{E} \left[\frac{\sigma^{2}
    (1)}{\widetilde{p}} + \frac{\sigma^{2} (0)}{1 - \widetilde{p}} \right] \geq \mathbb{E} \left[\frac{\sigma^{2}
    (1)}{1/2} + \frac{\sigma^{2} (0)}{1 - 1/2} \right]~.
\end{align*}
Define the variable $Z_m$:
\begin{equation*}
  Z_m = \frac{\widehat{\sigma}(1)}{\sigma(1)} \bigg/ \frac{\widehat{\sigma}(0)}{\sigma(0)}~.
\end{equation*}
We can then rewrite the above condition as
\begin{align*}
  & \mathbb{E} \left[ \left(1 + \frac{1}{Z_m}\frac{\sigma(0)}{\sigma(1)} \right) \sigma^2(1) + \left( 1 + Z_m\frac{\sigma(1)}{\sigma(0)}\right)\sigma^2(0)  \right] \geq 2\sigma^2(1) + 2\sigma^2(0) \\
  \Leftrightarrow  & \mathbb{E}\left[ \frac{1}{Z_m} + Z_m \right]\sigma(1)\sigma(0) \geq \sigma^2(1) + \sigma^2(0) \\
  \Leftrightarrow & \frac{\sigma^2(1)}{\sigma^2(0)} - \mathbb{E}\left[ \frac{1}{Z_m} + Z_m \right]\frac{\sigma(1)}{\sigma(0)} + 1 \leq 0~.
\end{align*}
By the quadratic formula, the above inequality is satisfied whenever
\begin{equation*}
  \frac{\sigma(1)}{\sigma(0)} \in \left[B_m - \sqrt{B_m^2 - 1 }, B_m + \sqrt{B_m^2 - 1} \right] = C_m ~,
\end{equation*}
where $B_m := \frac{1}{2} \mathbb{E}\left[ \frac{1}{Z_m} + Z_m \right]$. Note that
\begin{equation*}
  B_m^2 - (B_m^2 - 1) = 1 \Rightarrow \left(B_m - \sqrt{B_m^2 - 1 } \right)\left(B_m + \sqrt{B_m^2 - 1 } \right) = 1~,
\end{equation*}
so that
\begin{equation*}
  B_m - \sqrt{B_m^2 - 1 } = \frac{1}{B_m + \sqrt{B_m^2 - 1 }} = \frac{1}{x}~.
\end{equation*}
Next note that on $R_+$, $f(x) = \frac{1}{x} + x$ is strictly convex and attains its strict minimum at $x = 1$. Since $Z_m$ is non-degenerate, Jensen's inequality yields
\begin{equation*}
  2B_m > \frac{1}{\mathbb{E}\left[ Z_m\right]} + \mathbb{E}\left[ Z_m\right] \geq 2~.
\end{equation*}
Hence, $|C_m| > 0$. We can then write:
\begin{align*}
  |C_m| = 2\sqrt{B_m^2 - 1} = 2\sqrt{(B_m-1)^2 + 2(B_m-1)}~,
\end{align*}
where
\begin{equation*}
  B_m - 1 = \frac{1}{2} \left[\frac{\sigma(1)}{\sigma(0)} \, \text{Bias}\left(\widetilde{p}\right) +  \frac{\sigma(0)}{\sigma(1)} \, \text{Bias}\left(\frac{1}{\widetilde{p}}\right)\right] = W_m~.  \eqno\qed
\end{equation*}

\subsection{Proof of Proposition \ref{theorem--Cm_large_m}}

We start by evaluating the asymptotic distributions of $Z_m$. First,
\begin{align*}
  \sqrt{m}\left( \widehat{\sigma}^2(1) - \sigma^2(1) \right) & = \frac{1}{\sqrt{m}} \sum_{i = 1}^m \left(\left(Y_{i}(1) - \mu(1)\right)^2 - \sigma^2(1) \right) - \sqrt{n}\left(\bar{Y}(1) - \mu(1)\right)^2 + o_p(1) \\
  & \overset{d}{\to} \mathcal{N}\left( \, 0 \, , \mathbb{E}\left[ \left( Y_i(1) - \mu(1) \right)^4 \right] - \sigma^4(1) \right)~.
\end{align*}
By the Delta Method,
\begin{equation*}
  \sqrt{m}\left( \frac{\widehat{\sigma}^2(1)}{\sigma^2(1)} - 1 \right) \overset{d}{\to} \mathcal{N} \left( 0, \frac{\mathbb{E}\left[ \left( Y_i(1) - \mu(1) \right)^4 \right] - \sigma^4(1)}{4\sigma^4(1)} \right)~.
\end{equation*}
Similarly,
\begin{equation*}
  \sqrt{m}\left( \frac{\widehat{\sigma}^2(0)}{\sigma^2(0)} - 1 \right) \overset{d}{\to} \mathcal{N}\left( 0, \frac{\mathbb{E}\left[ \left( Y_i(0) - \mu(0) \right)^4 \right] - \sigma^4(0)}{4\sigma^4(0)} \right)~.
\end{equation*}
Since the above two displays contain independent random variables, another application of the Delta Method yields that
\begin{align*}
  \sqrt{m} \left(Z_m - 1\right) \overset{d}{\to} \mathcal{N}(0, V)~.
\end{align*}
Next, let $f(x) = x + 1/x$. Note that
\begin{align*}
  f'(x) = 1- \frac{1}{x^2} \quad , \quad & f'(1) = 0 ~,\\
  f''(x) = \frac{2}{x^3} \quad , \quad & f''(1) = 2~.
\end{align*}
By the second order Delta Method,
\begin{align*}
  m\left(\frac{1}{2}\left(\frac{1}{Z_m} + Z_m \right) - 1 \right) \overset{d}{\to} \frac{1}{2}V \cdot \chi^2_1~.
\end{align*}
Since the left hand side is an analytic function of sub-Gaussian random variables, all moments can be bounded uniformly in $m$. Conclude that:
\begin{equation}\label{equation--approx_B_m}
  m\left(B_m - 1\right) \to \frac{V}{2}~.
\end{equation}
Hence,
\begin{equation*}
  \sqrt{m} \left( \left(B_m + \sqrt{B_m^2 - 1}\right) - 1  \right) = \frac{m(B_m-1)}{\sqrt{m}} + \sqrt{m(B_m-1)\left(\frac{m(B_m-1)}{m}+2\right)} \to \sqrt{V}~. \eqno\qed
\end{equation*}

\subsection{Proof of Proposition \ref{theorem--eff_loss}}

By definition, $\tilde{p} = \left(1+ \frac{1}{Z_m}\frac{\sigma(0)}{\sigma(1)}\right)^{-1}$. Substituting this term into Definition \ref{definition--eff_loss} yields the expressions in terms of $B_m$. Equation \eqref{equation--approx_B_m} the proof of Proposition \ref{theorem--Cm_large_m} gives us the second set of expressions for when $m$ is large.

\subsection{Proof of Proposition \ref{theorem--regret}}

Recall from the proof of Proposition \ref{theorem--Cm_small_m} that $B_m > 1$.

We begin by showing that the displayed infimums are indeed lower bounds. Observe that
\begin{equation*}
  \mathcal{L}^d(F) \geq \left(\sigma(1) - \sigma(0)\right)^2 \geq - K~.
\end{equation*}
In the above display, the first inequality follows because $B_m > 1$ (see proof of Proposition \ref{theorem--Cm_small_m}). The second inequality follows by minimizing the middle term over $\mathcal{F}(K)$. That $\frac{1}{2}$ is a lower bound $\mathcal{L}^d(F)$ is immediate since $B_m > 0$. To see that $\mathcal{L}^d(F)$ and $\mathcal{L}^r(F)$ are attained, first fix the distribution of $\varepsilon(1) := Y(1) - \mu(1)$ and $\varepsilon(0) := Y(0) - \mu(0)$. Now choose some sequence $\omega_k \to 0$. Define $Y_k(1) = \mu(1) + \omega_k \varepsilon(1)$ and $Y_k(0) = Y(0)$ and denote their joint distribution $F_k$. Notice that each $F_k$ has the same $B_m(F_k) = \mathbb{E}_{F_k}[Z_m + \frac{1}{Z_m}]$. Then as $\omega_k \to 0$,
\begin{equation*}
  \mathcal{L}^d(F) \to -K \quad \mbox{ and } \quad \mathcal{L}^r(F) \to \frac{1}{2}~.
\end{equation*}

We next show that the variance of $\hat{\theta}_{\tilde{p}}$ is unbounded over the class of distributions with uniformly bounded $a^\text{th}$ moment for any $a$. We now construct a simple counterexample in which we essentially force division by $0$. For a given $a$, let $Y(0) \sim N(0,1)$. Let $Y(1) \sim X + \omega W$, where $W \sim N(0,1)$, $W \indep \, X$ and
\begin{align*}
  X = \begin{cases}
  - 1& \mbox{ w.p. } \frac{1}{2\kappa} \\
  0 & \mbox{ w.p. } 1 - \frac{1}{\kappa} \\
  1 & \mbox{ w.p. } \frac{1}{2\kappa}
  \end{cases}~.
\end{align*}
By construction, $E[Y(1)^p] \leq C_p \left(X^p + W^p \right) \leq K$ (we can scale $Y(1)$ if necessary). Now, let $A$ be the event that $X_1 = X_2 = ... = X_m = 0$. Then,
\begin{align*}
  B_m & \geq E\left[Z_m + \frac{1}{Z_m} \; \big\lvert A \right] P(A)  \\
  & = \left(\omega + \frac{1}{\omega}\right)\mathbb{E}\left[F_{\frac{m}{2}, \frac{m}{2}}\right]\left(1-\frac{1}{\kappa}\right)^m \\
  & \to \infty \mbox{ as } \omega \to 0~.
\end{align*}
To show non-uniformity in the large pilot regime, we can set $\kappa = m$ and $\omega = 1/m$. Then $B_m \approx m e^{-1} \to \infty$.

\section{Additional Empirical Examples}
\label{section--empirical_additional}

We revisit the first 10 completed RCTs in the AER RCT Registry. This section
contains the empirical examples omitted from the main text. They are:

\begin{itemize}
\item Section \ref{subsection--empirical--dillon}: \citet{dillon2017cognitive}
\item Section \ref{subsection--empirical--finkelstein}:
  \citet{finkelstein2012oregon}
\item Section \ref{subsection--empirical--mckenzie}:
  \citet{mckenzie2017identifying}
\item Section \ref{subsection--empirical--chong}: \citet{chong2015ineffective}
\item Section \ref{subsection--empirical--deming}:
  \citet{2016demingValuePostsecondaryCredentials}
\item Section \ref{subsection--empirical--bloom}:
  \citet{2014bloomDoesWorkingHome}
\item Section \ref{subsection--empirical--galiani}:
  \citet{2013galianiHeterogeneousImpactConditional}
\item Section \ref{subsection--empirical--bryan}: \citet{2015bryanReferrals}
\end{itemize}

\subsection{\citet{dillon2017cognitive}}
\label{subsection--empirical--dillon}

\citet{dillon2017cognitive} conduct an RCT to test the hypothesis that math game play in pre-school prepares poor children for formal math in primary school. Their study, conducted in Delhi, India, with the organization Pratham, involved 214 pre-schools with 1540 children, with treatment assigned at the school level. In the Math treatment arm, children were led by facilitators to play math games over the course of four months, while the control group received lessons according to Pratham's usual curriculum. To distinguish the effect of the math games from the effect of engagement with adults, the experiment further involved a Social treatment arm, where social games were played. The outcomes of interest are Math Skills -- subdivided into Symbolic Math Skills and Non-Symbolic Math Skills -- as well as Social Skills, as measured by Pratham's standardized tests. Since the authors are interested in persistence of treatment effects, they measure these outcomes at the following times after intervention: 0-3 months (Endline 1), 6-9 months (Endline 2) and 12-15 months (Endline 3). They find that the Math intervention has positive effects on Non-Symbolic Mathematical skills across all three Endlines, while Symbolic Mathematical skills only improves in Endline 1.

Table \ref{table:empirical--dillon--indiv} displays the standard deviation of each outcome variable by treatment arm, computed at the individual level. In the absence of correlation among students in the same pre-school, and assuming that treatment is assigned at the individual level, the numbers shown are the relevant empirical counterparts to $\sigma(1)$ and $\sigma(0)$. We see that the outcomes are relatively homoskedastic across the outcome variables and across time. The ratio $\sigma(1)\sigma(0)$ falls between 0.94 and 1.31, suggesting that naive experiment will do well when pilots are small.

\begin{table}[htbp]
  \centering
  \caption{Individual Level Heteroskedasticity in \citet{dillon2017cognitive}}
  \begin{tabular}{clccccc}
    \hline\hline
    Endline & Outcome & Math  & Social & Control & Math/Control & Social/Control \\
    \midrule
    \multirow{4}[0]{*}{1} & Math & 0.73  & 0.68  & 0.69  & 1.06  & 0.99 \\
    & Symbolic Math & 0.74  & 0.78  & 0.77  & 0.96  & 1.01 \\
    & Non-Symbolic Math & 0.94  & 0.81  & 0.77  & 1.21  & 1.04 \\
    & Social & 1.18  & 1.41  & 1.07  & 1.10  & 1.31 \\
    \midrule
    \multirow{4}[0]{*}{2} & Math & 0.71  & 0.73  & 0.69  & 1.04  & 1.06 \\
    & Symbolic Math & 0.72  & 0.74  & 0.71  & 1.02  & 1.05 \\
    & Non-Symbolic Math & 0.98  & 0.99  & 0.92  & 1.06  & 1.07 \\
    & Social & 0.92  & 0.96  & 0.99  & 0.94  & 0.97 \\
    \midrule
    \multirow{4}[0]{*}{3} & Math & 0.78  & 0.70  & 0.75  & 1.04  & 0.93 \\
    & Symbolic Math & 0.72  & 0.68  & 0.74  & 0.98  & 0.92 \\
    & Non-Symbolic Math & 1.17  & 1.09  & 1.06  & 1.11  & 1.03 \\
    & Social & 1.01  & 1.05  & 1.07  & 0.94  & 0.98 \\
    \hline\hline
  \end{tabular}%
  \label{table:empirical--dillon--indiv}%
\end{table}%

Suppose we are concerned about correlation across students in the same pre-school. We can redefine our unit of observation to be the school by taking averages across students in the same school. The Neyman Allocation then tells us how many schools to allocate to treatment. In this case, the standard deviation in the mean across schools is the relevant counterpart to $\sigma(1)$ and $\sigma(0)$. They are presented in Table \ref{table:empirical--dillon--clust}. As before, we see that outcomes are relatively homoskedastic across schools, though less so than in the individual level case. Nonetheless, the ratio falls between $0.84$ and $1.55$. Once we consider schools to be the unit of treatment, however, the effective pilot size also shrinks, such that the drawbacks of the estimated Neyman Allocation may be even more pronounced. All in all, the \citet{dillon2017cognitive} example supports our case of relative homoskedasticity in empirical applications.

\begin{table}[htbp]
  \centering\small
  \caption{School Level Heteroskedasticity in \citet{dillon2017cognitive}}
  \begin{tabular}{clccccc}
    \hline\hline
    Endline & Outcome & Math  & Social & Control & Math/Control & Social/Control \\
    \midrule
    \multirow{4}[0]{*}{1} & Math & 0.41  & 0.38  & 0.30  & 1.34  & 1.24 \\
    & Symbolic Math & 0.39  & 0.42  & 0.34  & 1.14  & 1.22 \\
    & Non-Symbolic Math & 0.51  & 0.40  & 0.33  & 1.55  & 1.22 \\
    & All Social & 0.51  & 0.71  & 0.46  & 1.10  & 1.55 \\
    \midrule
    \multirow{4}[0]{*}{2} & All Math & 0.39  & 0.35  & 0.36  & 1.10  & 0.97 \\
    & Symbolic Math & 0.40  & 0.35  & 0.36  & 1.11  & 0.96 \\
    & Non-Symbolic Math & 0.46  & 0.42  & 0.43  & 1.06  & 0.98 \\
    & Social & 0.41  & 0.48  & 0.49  & 0.84  & 0.97 \\
    \midrule
    \multirow{4}[0]{*}{3} & All Math & 0.44  & 0.40  & 0.39  & 1.15  & 1.04 \\
    & Symbolic Math & 0.40  & 0.35  & 0.37  & 1.06  & 0.94 \\
    & Non-Symbolic Math & 0.65  & 0.63  & 0.53  & 1.23  & 1.20 \\
    & Social & 0.55  & 0.59  & 0.50  & 1.10  & 1.20 \\
    \hline\hline
  \end{tabular}%
  \label{table:empirical--dillon--clust}%
\end{table}%

\subsection{\citet{finkelstein2012oregon}}
\label{subsection--empirical--finkelstein}

\citet{finkelstein2012oregon} study the Oregon Health Insurance Experiment, in which uninsured, low-income adults were randomly given the opportunity to apply for Medicaid. Over the course of a month in February 2008, Oregon conducted extensive public awareness campaign to encourage participation in the lottery. From a total of 89,824 sign-ups, 35,169 individuals (from 29,664 households) were selected. They, and any members of their households were then given the opportunity to apply for Medicaid. Hence, treatment occurred at the household level.

The authors used the data to study a variety of outcomes. In this section, we focus on their first set of results, which concern healthcare utilization. In particular, we revisit the outcome variables used in Tables V and VI of \citet{finkelstein2012oregon}, which are obtained from survey data (as opposed to administrative data), and are hence publicly available. Inline with the authors' results on the Intent-to-Treat effect, we define the treated group as those selected by the lottery. We note that the authors apply sampling weights to correct for differential response rates to the survey. We follow their weighting scheme in computing our results.

The standard deviation of individual-level outcome are presented in Table \ref{table:empirical--finkelstein--indiv}. Results taking household to be the unit of observation are presented in Table \ref{table:empirical--finkelstein--clust}. Across both tables, we see that that the standard deviations in outcomes are remarkably similar across treatment and control groups. They are also very similar across individual and household level groups, since households with more than one person represented less than 5\% of the survey sample.

\begin{table}[htbp]
  \centering\footnotesize
  \caption{Individual Level Heteroskedasticity in \citet{finkelstein2012oregon}}
  \begin{tabular}{clccc}\hline\hline
    & Outcome & \multicolumn{1}{l}{Treatment} & \multicolumn{1}{l}{Control} & \multicolumn{1}{l}{Treat./Cont.} \\
    \midrule
    \multirow{4}[0]{*}{\shortstack{Extensive\\Margin}} & Prescription drugs currently & 0.48  & 0.48  & 0.99 \\
    & Outpatient visits last six months & 0.48  & 0.49  & 0.98 \\
    & ER visits last six months & 0.44  & 0.44  & 1.00 \\
    & Inpatient hospital admissions last six months & 0.26  & 0.26  & 1.00 \\
    \midrule
    \multirow{4}[0]{*}{\shortstack{Total\\Utilization}} & Prescription drugs currently & 2.90  & 2.88  & 1.01 \\
    & Outpatient visits last six months & 3.29  & 3.09  & 1.07 \\
    & ER visits last six months & 1.01  & 1.04  & 0.97 \\
    & Inpatient hospital admissions last six months & 0.42  & 0.40  & 1.04 \\
    \midrule
    \multirow{4}[0]{*}{\shortstack{Preventative\\Care}} & Blood cholesterol checked (ever) & 0.48  & 0.48  & 0.98 \\
    & Blood tested for high blood sugar (ever) & 0.48  & 0.49  & 0.99 \\
    & Mammogram within last 12 months (women $\geq$ 40) & 0.48  & 0.46  & 1.04 \\
    & Pap test within last 12 months (women) & 0.50  & 0.49  & 1.01 \\
    \hline\hline
  \end{tabular}%
  \label{table:empirical--finkelstein--indiv}%
\end{table}%

\begin{table}[htbp]
  \centering\footnotesize
  \caption{Household Level Heteroskedasticity in \citet{finkelstein2012oregon}}
  \begin{tabular}{clccc}\hline\hline
    & Outcome & \multicolumn{1}{l}{Treatment} & \multicolumn{1}{l}{Control} & \multicolumn{1}{l}{Treat./Cont.} \\
    \midrule
    \multirow{4}[0]{*}{\shortstack{Extensive\\Margin}} & Prescription drugs currently & 0.46  & 0.47  & 0.99 \\
    & Outpatient visits last six months & 0.47  & 0.48  & 0.97 \\
    & ER visits last six months & 0.43  & 0.44  & 0.99 \\
    & Inpatient hospital admissions last six months & 0.26  & 0.26  & 0.99 \\
    \midrule
    \multirow{4}[0]{*}{\shortstack{Total\\Utilization}} & Prescription drugs currently & 2.88  & 2.86  & 1.01 \\
    & Outpatient visits last six months & 3.30  & 3.09  & 1.07 \\
    & ER visits last six months & 1.01  & 1.04  & 0.98 \\
    & Inpatient hospital admissions last six months & 0.41  & 0.40  & 1.02 \\
    \midrule
    \multirow{4}[0]{*}{\shortstack{Preventative\\Care}} & Blood cholesterol checked (ever) & 0.46  & 0.48  & 0.97 \\
    & Blood tested for high blood sugar (ever) & 0.47  & 0.48  & 0.98 \\
    & Mammogram within last 12 months (women $\geq$ 40) & 0.48  & 0.46  & 1.04 \\
    & Pap test within last 12 months (women) & 0.50  & 0.49  & 1.01 \\
    \hline\hline
  \end{tabular}%
  \label{table:empirical--finkelstein--clust}%
\end{table}%

\subsection{\citet{mckenzie2017identifying}}
\label{subsection--empirical--mckenzie}

Business plan competitions are growing in popularity as a way of fostering high growth entrepreneurship in developing countries. \citet{mckenzie2017identifying} studies the Youth Enterprise With Innovation in Nigeria (YouWiN!) program, which distributed up to US\$64,000 to winners. A portion of the awards were reserved for business plans that were clearly superior to the rest. 1,841 entrepreneurs, determined to be of medium quality, were entered into a lottery, from which 729 were selected for the award. Three rounds of surveys were then conducted at 1,2 and 3 years after the application respectively.

We focus on the first set of results in \citet{mckenzie2017identifying} -- presented in Table 2 -- concerning the effect of the award on start-up and survival. Here, they find that the grant persistently increased the probability that the entrepreneur was operating a firm, as well as the number of hours they spent in self-employment. We present the standard deviations of these outcome variables in Table \ref{table:empirical--mckenzie--sd}. Here we see that hours in self employment is roughly homoskedastic across all three periods. However, the outcome on whether the entrepreneur is operating a firm is arguably highly heteroskedastic, with standard deviations that is as small as 0.44 that of the control group. As in our earlier example, we find high kurtosis in these outcome variables, displayed in Table \ref{table:empirical--mckenzie--kurtosis}.

We do not estimate $C_m$ in this example because almost all entrepreneurs in the treated group operate their own firms in the sample. As such, a small random sub-sample (e.g. of size below 200) from this group has variance 0 with high probability, impeding the estimation of $C_m$. These pathological cases are revealing. In a pilot, if the treated group has variance $0$ in the outcome, the estimated Neyman Allocation assigns $0$ units to treatment in the full experiment. The high probability of such an ``extreme" outcome with small pilots is precisely the danger which we are warning against. We conclude that the estimated Neyman from a small pilot will likely lead to adverse results given the DGP in \citet{mckenzie2017identifying}.

\begin{table}[htbp]
  \centering\small
  \caption{Heteroskedasticity in \citet{mckenzie2017identifying}}
    \begin{tabular}{lcccccccc}\hline\hline
    \multicolumn{1}{c}{\multirow{2}[1]{*}{Outcome}} &       & \multicolumn{3}{c}{New Firms} &       & \multicolumn{3}{c}{Existing Firms} \\
          &       & Treat. & Cont. & Ratio &       & Treat. & Cont. & Ratio \\
\cmidrule{1-1}\cmidrule{3-5}\cmidrule{7-9}    Operates a Firm at Round 1 &       & 0.43  & 0.50  & 0.86  &       & 0.21  & 0.34  & 0.63 \\
    Operates a Firm at Round 2 &       & 0.27  & 0.50  & 0.55  &       & 0.16  & 0.36  & 0.44 \\
    Operates a Firm at Round 3 &       & 0.28  & 0.50  & 0.56  &       & 0.20  & 0.43  & 0.48 \\
    Weekly Hours of Self Emp. at Round 1 &       & 29.40 & 29.75 & 0.99  &       & 25.74 & 27.81 & 0.93 \\
    Weekly Hours of Self Emp. at Round 2 &       & 24.98 & 28.62 & 0.87  &       & 24.51 & 29.71 & 0.82 \\
    Weekly Hours of Self Emp. at Round 3 &       & 24.77 & 25.85 & 0.96  &       & 25.09 & 26.10 & 0.96 \\
    \hline\hline
    \end{tabular}%
  \label{table:empirical--mckenzie--sd}%
\end{table}%

\begin{table}[htbp]
  \centering\small
  \caption{Kurtosis in \citet{mckenzie2017identifying}}
    \begin{tabular}{lcccccc}\hline\hline
    \multicolumn{1}{c}{\multirow{2}[1]{*}{Outcome}} &       & \multicolumn{2}{c}{New Firms} &       & \multicolumn{2}{c}{Existing Firms} \\
          &       & Treat. & Cont. &       & Treat. & Cont. \\
\cmidrule{1-1}\cmidrule{3-4}\cmidrule{6-7}    Operates a Firm at Round 1 &       & 2.52  & 1.04  &       & 19.68 & 5.91 \\
    Operates a Firm at Round 2 &       & 10.69 & 1.08  &       & 35.46 & 4.58 \\
    Operates a Firm at Round 3 &       & 9.62  & 1.03  &       & 21.05 & 2.47 \\
    Weekly Hours of Self Emp. at Round 1 &       & 1.99  & 3.08  &       & 3.93  & 3.08 \\
    Weekly Hours of Self Emp. at Round 2 &       & 2.77  & 2.92  &       & 3.03  & 2.42 \\
    Weekly Hours of Self Emp. at Round 3 &       & 2.09  & 3.28  &       & 3.47  & 2.15 \\
    \hline\hline
    \end{tabular}%
  \label{table:empirical--mckenzie--kurtosis}%
\end{table}%

\subsection{\citet{chong2015ineffective}}
\label{subsection--empirical--chong}

Partnering with the Peruvian nongovernmental organization PRISMA, \citet{chong2015ineffective} conducted two RCTs to investigate the efficacy of various interventions in encouraging recycling. First, the Participation Study considers the following $9$ different messaging strategies and their relative success in enrolling members into recycling programs:
\begin{enumerate}
  \item Norms: Rich and Poor. Norm messaging focus on communicating high recycling rate of either a rich or poor reference neighborhoods, encouraging conformity.
  \item Signal: Rich, Poor and Local. Signal messaging informs the targets that their recycling behavior will be known to either a nearby neighborhood (Local), a distal neighborhood of varying wealth (Rich or Poor), affecting the targets reputation.
  \item Authority: Religious or Municipal. Authority messaging communicates that a higher authority, either religious or local governmental, advocates recycling.
  \item Information: Environmental or Social. Informational messaging communicated the benefits of recycling, either to the environment or to the local society (e.g. by creating jobs).
\end{enumerate}
Out of a total of 6,718 households, approximately 600 were assigned to each treatment arm, with the exception of Signal: Local, which were assigned 932 participants. 1,157 households were assigned to the control group. Three measures of participation were considered:
\begin{enumerate}
  \item ``Participates any time" is an indicator that takes the value 1 if a household turned in residuals over the course of the study.
  \item ``Participation Ratio" is the number of times a household turns in residual over the total number of opportunities they had to turn in residuals.
  \item  ``Participates during either of last two visits" is an indicator that takes value 1 if the household turned in residual during one of the last two canvassing weeks.
\end{enumerate}
The results in Table 3 of \citet{chong2015ineffective} shows that messaging had no effect in increasing participation in the program. Table \ref{table:empirical--chong--sd} displays the standard deviation of the various outcomes by treatment type. Table \ref{table:empirical--chong--ratios}, shows the ratio of the standard deviation in the outcome variable of each treatment group, with respect to that of the control group. Here, we see that the outcomes are highly homoskedastic, suggesting little scope for improvement over the naive experiment.

\begin{table}[htbp]
  \centering\footnotesize
  \caption{S.D. of outcome by treatment type in the Participation Study. See text for definitions of outcome and treatment.}
    \begin{tabular}{cccccccccccccccc}\hline\hline
    \multicolumn{1}{c}{\multirow{2}[1]{*}{Outcome}} &       & \multirow{2}[1]{*}{Control} &       & \multicolumn{2}{c}{Norms} &       & \multicolumn{3}{c}{Signal} &       & \multicolumn{2}{c}{Authority} &       & \multicolumn{2}{c}{Info.} \\
          &       &       &       & Rich  & Poor  &       & Rich  & Poor  & Local &       & Reli. & Muni. &       & Env.  & Social \\
\cmidrule{1-1}\cmidrule{3-3}\cmidrule{5-6}\cmidrule{8-10}\cmidrule{12-13}\cmidrule{15-16}    1     &       & 0.500 &       & 0.500 & 0.500 &       & 0.500 & 0.500 & 0.500 &       & 0.500 & 0.500 &       & 0.500 & 0.500 \\
    2     &       & 0.389 &       & 0.389 & 0.392 &       & 0.403 & 0.385 & 0.393 &       & 0.397 & 0.388 &       & 0.396 & 0.392 \\
    3     &       & 0.490 &       & 0.486 & 0.494 &       & 0.493 & 0.486 & 0.489 &       & 0.491 & 0.491 &       & 0.494 & 0.493 \\
    \hline\hline
    \end{tabular}%
  \label{table:empirical--chong--sd}%
\end{table}%

\begin{table}[htbp]
  \centering\small
  \caption{Ratio of the S.D w.r.t. the control group in the Participation Study.}
    \begin{tabular}{cccccccccccccc}\hline\hline
    \multirow{2}[1]{*}{Outcome} &       & Norms &       &       & Signal &       &       &       & Authority &       &       & \multicolumn{2}{c}{Info.} \\
          &       & Rich  & Poor  &       & Rich  & Poor  & Local &       & Reli. & Muni. &       & Env.  & Social \\
\cmidrule{1-1}\cmidrule{3-4}\cmidrule{6-8}\cmidrule{10-11}\cmidrule{13-14}    1     &       & 1.000 & 0.999 &       & 1.000 & 1.000 & 1.000 &       & 1.000 & 0.999 &       & 1.000 & 1.000 \\
    2     &       & 1.000 & 1.008 &       & 1.036 & 0.989 & 1.012 &       & 1.022 & 0.998 &       & 1.018 & 1.008 \\
    3     &       & 0.992 & 1.007 &       & 1.005 & 0.990 & 0.997 &       & 1.000 & 1.001 &       & 1.007 & 1.005 \\
    \hline\hline
    \end{tabular}%
  \label{table:empirical--chong--ratios}%
\end{table}%

The second experiment is the Participation Intensity Study. The outcomes of interest are the following measures recycling intensity:
\begin{enumerate}
  \item Percentage of visits in which household turned in residuals
  \item Average number of bins turned in per week
  \item Average weight (in kg) of recyclables turned in per week
  \item Average market value of recyclables given per week
  \item Average percentage of contamination (non-recyclables mixed into recycling) per week.
\end{enumerate}
The treatments of interest are (1) providing recycling bins to households and (2) sending SMS reminders for recycling.\footnote{The authors also consider providing bins with and without instructions as well as generic vs personalized SMSes. However, these finer definitions leads to treatment arms with fewer 50 households. Hence we focus on the coarser definition of the treatments, as employed in panel 4A.} Of the 1,781 households in this study, 182 were received Bin and SMS. 417 received the Bin only treatment. 369 received the SMS only treatment, leaving 817 in the control group. The authors find, in Table 4A that bin provision was highly effective in increasing recycling, though SMS reminders had no effect. We compute the standard deviation in each of their outcome variables by treatment type in Table \ref{table:empirical--chong--intensity_ratios}. Also displayed is the ratio of these standard deviation with respect to the control group. Again, we see strong evidence of homoskedasticity, suggesting that the naive experiment will perform well in this scenario.

\begin{table}[htbp]
  \centering\footnotesize
  \caption{S.D. of outcome by treatment type in the Participation Intensity Study. See text for definitions of outcome and treatment.}
    \begin{tabular}{cccccccccc}\hline\hline
    \multirow{2}[1]{*}{Outcome} &       & \multicolumn{4}{c}{Standard Deviation} &       & \multicolumn{3}{c}{Ratio of S.D. w.r.t. Control} \\
          &       & Control & SMS only & Bin only & SMS \& Bin &       & SMS only & Bin only & SMS \& Bin \\
\cmidrule{1-1}\cmidrule{3-6}\cmidrule{8-10}    1     &       & 0.262 & 0.286 & 0.233 & 0.227 &       & 1.092 & 0.891 & 0.867 \\
    2     &       & 0.404 & 0.371 & 0.441 & 0.375 &       & 0.919 & 1.092 & 0.927 \\
    3     &       & 0.744 & 0.646 & 0.756 & 0.727 &       & 0.869 & 1.017 & 0.978 \\
    4     &       & 0.418 & 0.371 & 0.416 & 0.399 &       & 0.889 & 0.996 & 0.955 \\
    5     &       & 0.156 & 0.145 & 0.136 & 0.128 &       & 0.928 & 0.870 & 0.822 \\
    \hline\hline
    \end{tabular}%
  \label{table:empirical--chong--intensity_ratios}%
\end{table}%

\subsection{\citet{2016demingValuePostsecondaryCredentials}}
\label{subsection--empirical--deming}

\citet{2016demingValuePostsecondaryCredentials} study employers' perceptions of
the value of post-secondary degrees using a field experiment. The experimental
units are fictitious resumes to be used in applications to vacancies posted on a
large online job board. Their focus is on degrees and certificates awarded in
the two largest occupational categories in the United States: business and
health. Resumes are randomly assigned sector and selectivity of
(degree-awarding) institutions. Fictitious resumes are created using a vast
online database of actual resumes of job seekers, with applicant characteristics
varying randomly (i.e. characteristics are randomly assigned). Outcomes are
callback rates. There are three main comparisons in the paper:
\begin{itemize}
  \item for-profit vs. public institutions,
  \item for-profits that are online vs. brick-and-mortar (with a local
  presence),
  \item more selective vs. less selective public institutions.
\end{itemize}
\citet{2016demingValuePostsecondaryCredentials} find that BA degrees in business
from large online for-profit institutions are more 22\% less likely to receive a
callback than applicants with similar degrees from non-selective public schools
when the job vacancy requires a BA. When a business job opening does not list a
BA requirement, they find no significant overall advantage to having a
post-secondary degree. For health jobs, resumes with certificates from
for-profit institutions are 57\% less likely to receive a callback than those
with similar certificates from public institutions when the job listing does not
require a post-secondary certificate. No significant difference in callback
rates are found when the health job listing requires a certificate.

Table \ref{table--deming} reports standard deviations across treatment arms
across the various sub-populations of interest in
\citet{2016demingValuePostsecondaryCredentials}. Since in most sub-populations,
there are more than two treatment arms, we do not report ratios. However,
pairwise comparisons between treatment arms within any chosen subpopulation
shows strong evidence of relative homoskedasticity across the board.

\begin{table}[htbp]
\caption{Performance impact outcomes: standard deviations}
\begin{center}
\label{table--deming}
\begin{tabular}{llc}
\hline\hline
Experimental Population               & Treatment Arm               & S.D. \\ \hline
Business jobs without BA requirements & No degree                   & 0.3054              \\
                                      & AA (for profit)             & 0.3026              \\
                                      & AA (public)                 & 0.3053              \\
                                      & BA (for profit)             & 0.3071              \\ \hline
Business jobs with BA requirements    & BA (for profit, online)     & 0.2522              \\
                                      & BA (for profit, not online) & 0.2209              \\
                                      & BA (public, selective)      & 0.2879              \\
                                      & BA (public, not selective)  & 0.2595              \\ \hline
Health job without cert. requirement  & No certificate              & 0.2900              \\
                                      & Certificate (for profit)    & 0.2922              \\
                                      & Certificate (public)        & 0.3014              \\ \hline
Health job with cert. requirement     & Certificate (for profit)    & 0.2400              \\
                                      & Certificate (public)        & 0.2681              \\
                                      \hline\hline
\end{tabular}
\end{center}
\end{table}

\subsection{\citet{2014bloomDoesWorkingHome}}
\label{subsection--empirical--bloom}

\citet{2014bloomDoesWorkingHome} study the effect of working from home on
employees' productivity via a randomized experiment at Ctrip, a NASDAQ-listed
Chinese travel agency with 16000 employees. The main concern is whether or not
working from leads to shirking. Ctrip decided to run a nine-month experiment on
working from home. They asked the 996 employees in the airfare and hotel
departments of the Shanghai call center if they were interested in working from
home four days a week and one day in the office. 503 of these employees were
interested and of these, 249 were qualified to take part on the basis of tenure,
broadband access and access to private work space at home. Qualified employees
were assigned to working from home if they had even-numbered birthdays so that
those with odd-numbered birthdays formed the control group. The treatment and
control groups were comprised of 131 and 118 employees respectively. The only
difference between the two groups was location of work -- both groups used the
same equipment, faced the same workload and were compensated under the same pay
system. The authors find a 13\% increase in productivity of which the main
source of improvement was a 9\% increase in the number of minutes worked during
a shift. The remaining 4\% came from an increase in the number of calls per
minute worked.

Table \ref{table--bloom} reports standard deviations for treatment and control
groups in the experiment as well as the standard deviation ratios for the main
outcomes of interest in \citet{2014bloomDoesWorkingHome}. Strong evidence of
relative homoskedasticity with respect to treatment status presents in this
study as well -- suggesting that the balanced allocation would outperform the
FNA.

\begin{table}[htbp]
\caption{Performance impact outcomes: standard deviations}
\begin{center}
\label{table--bloom}
\begin{tabular}{lccc}
\hline\hline
Outcome Variables   & Control S.D. & Treated S.D. & Ratio  \\ \hline
Overall Performance & 1.0049       & 1.0035       & 0.9986 \\
Phone calls         & 0.9775       & 0.7502       & 0.7675 \\
Log phone calls     & 0.2476       & 0.1764       & 0.7123 \\
Log call per sec    & 0.0217       & 0.0299       & 1.3786 \\
Log call length     & 0.2701       & 0.2729       & 1.0105 \\
\hline\hline
\end{tabular}
\end{center}
\end{table}

\subsection{\citet{2013galianiHeterogeneousImpactConditional}}
\label{subsection--empirical--galiani}

\citet{2013galianiHeterogeneousImpactConditional} use the Honduran PRAF
experiment to study the impact of conditional cash transfers (CCT) on the
likelihood of children to work versus enrolling in school. The PRAF experiment
randomly allocated CCT's among 70 municipalities. These 70 were chosen out of a
total of 298 on the basis of mean heights-for-age z-scores of first graders. The
70 municipalities were further assigned to four treatment arms termed G1, G2,
G3, G4. G1 received CCT's in education and health. G2 received CCT's in
addition to direct investment in education and health centers. G3 received only
direct investments and finally, G4 served as the control group and received no
interventions. The 70 municipalities were further divided into 5 strata each
consisting of 14 municipalities on the basis of quintiles of mean
height-for-age. Random assignment was performed within these strata (stratified
randomization). The final sample consisted of 20 municipalities in G1, 20 in
G2, 10 in G3, and 20 in G4. The authors match the experimental data with census
data and use the latter to construct the outcomes of interest which are three
dummy variables. The first is an indicator for whether a child is enrolled in
and attending school during the time of the census. The second indicates whether
the child worked during the week prior to the census or conditional on a
negative response to the former, whether they reported non-wage employment in a
family farm or business. The third indicates whether the child worked
exclusively on household chores. The authors find that overall, children
eligible for CCTs were 8\% more likely to enroll in school and 3\% less likely
to work.

Table \ref{table--galiani} reports standard deviations for the outcomes of
interest across treatment arms. Again, for a given outcome, pairwise comparisons
across arms yields evidence of relative homoskedasticity with respect to
treatment status. We conclude that in this case, the balanced allocation would
outperform the FNA.

\begin{table}[htbp]
\caption{St. dev. across treatment groups G1, G2, G3, G4}
\begin{center}
\label{table--galiani}
\begin{tabular}{lcccc}
\hline\hline
Outcome            & G1     & G2     & G3     & G4     \\ \hline
Enrolled in school & 0.4393 & 0.4474 & 0.4812 & 0.4769 \\
Works outside home & 0.2637 & 0.2267 & 0.2893 & 0.2986 \\
Works only in home & 0.3015 & 0.2841 & 0.3478 & 0.3409 \\
\hline\hline
\end{tabular}
\end{center}
\end{table}

\subsection{\citet{2015bryanReferrals}}
\label{subsection--empirical--bryan}

\citet{2015bryanReferrals} conduct a field experiment to study efficacy of peer
intermediation in mitigating adverse selection and moral hazard in credit
markets. To identify the effects of peer screening and enforcement, they use a
two-stage referral incentive field experiment. The experiment was conducted
through Opportunity Finance South Africa (Opportunity), a for-profit lender in
the consumer micro-loan market. Over the period of February 2008 through July
2009, Opportunity offered individuals approved for a loan the option to
participate in its ``Refer-A-Friend'' program. Referred individuals earned R40
if they brought in a referral card and were approved for a loan. The referrer
could earn R100 for referring someone who was subsequently approved for and/or
repaid the loan, depending on the referrer's incentive contract. Referrers were
randomly assigned to one of two ex-ante incentive contracts:
\begin{itemize}
  \item Approval incentives: the referrer would be paid only if the referred was
  approved for a loan.
  \item Repayment incentive: the referrer would be paid only if the referred
  successfully repaid the loan.
\end{itemize}
Among referrers whose referred friends were approved for a loan, Opportunity
randomly selected half to be surprised with an ex-post incentive change:
\begin{itemize}
  \item Half among the ex-ante approval group were phoned and told that in
  addition to the R100 approval bonus, they would receive an additional R100 if
  the loan was successfully repaid by the referrer.
  \item Half among the ex-ante repayment group were phoned and told that they
  would receive the R100 now, and that receipt of the bonus would no longer be
  conditional on repayment of the loan by the referrer.
\end{itemize}
The overall incentive structure is as follows
\begin{itemize}
  \item Ex-ante and ex-post approval (EA = A): no enforcement or screening
  incentive.
  \item Ex-ante repayment and ex-post approval (EA = R): screening
  incentive.
  \item Ex-ante approval and ex-post repayment (EA = A, EP = R): Enforcement
  incentive.
  \item Ex-ante repayment and ex-post repayment (EA = R, EP = R): Enforcement
  and screening incentive.
\end{itemize}
The authors find no evidence of screening but do find large enforcement
effects.

Table \ref{table--bryan} reports standard deviations in the main outcomes of
interest in \citet{2015bryanReferrals} across the four treatment arms. For a
given outcome, pairwise comparisons across arms yields evidence of relative
homoskedasticity with respect to treatment status. We conclude that in this
case, the balanced allocation would outperform the FNA.

\begin{table}[htbp]
\caption{St. dev. across treatment groups G1, G2, G3, G4}
\begin{center}
\label{table--bryan}
\begin{tabular}{lcccc}
\hline\hline
\shortstack{Outcome \\ \hfill}                               & \shortstack{EA = A \\ \hfill} & \shortstack{EA = A, \\ EP = R} & \shortstack{EA = R \\ \hfill} & \shortstack{EA = R, \\ EP = R} \\ \hline
Penalty interest                      & 0.4919 & 0.4407         & 0.4488 & 0.5043         \\
Positive balance owing at maturity    & 0.4086 & 0.2959         & 0.3613 & 0.4225         \\
Proportion of value owing at maturity & 0.4088 & 0.3106         & 0.3153 & 0.5526         \\
Loan charged off                      & 0.3652 & 0.2147         & 0.2917 & 0.3950         \\ \hline\hline
\end{tabular}
\end{center}
\end{table}

\clearpage
\newpage

\section{Additional Simulations}
\vspace{-7mm}
\subsection{MSE for $m = 50$}\label{appendix--MSE}

\begin{figure}[h!]
\vspace{-3mm}
\centering
\includegraphics[width=0.66\linewidth]{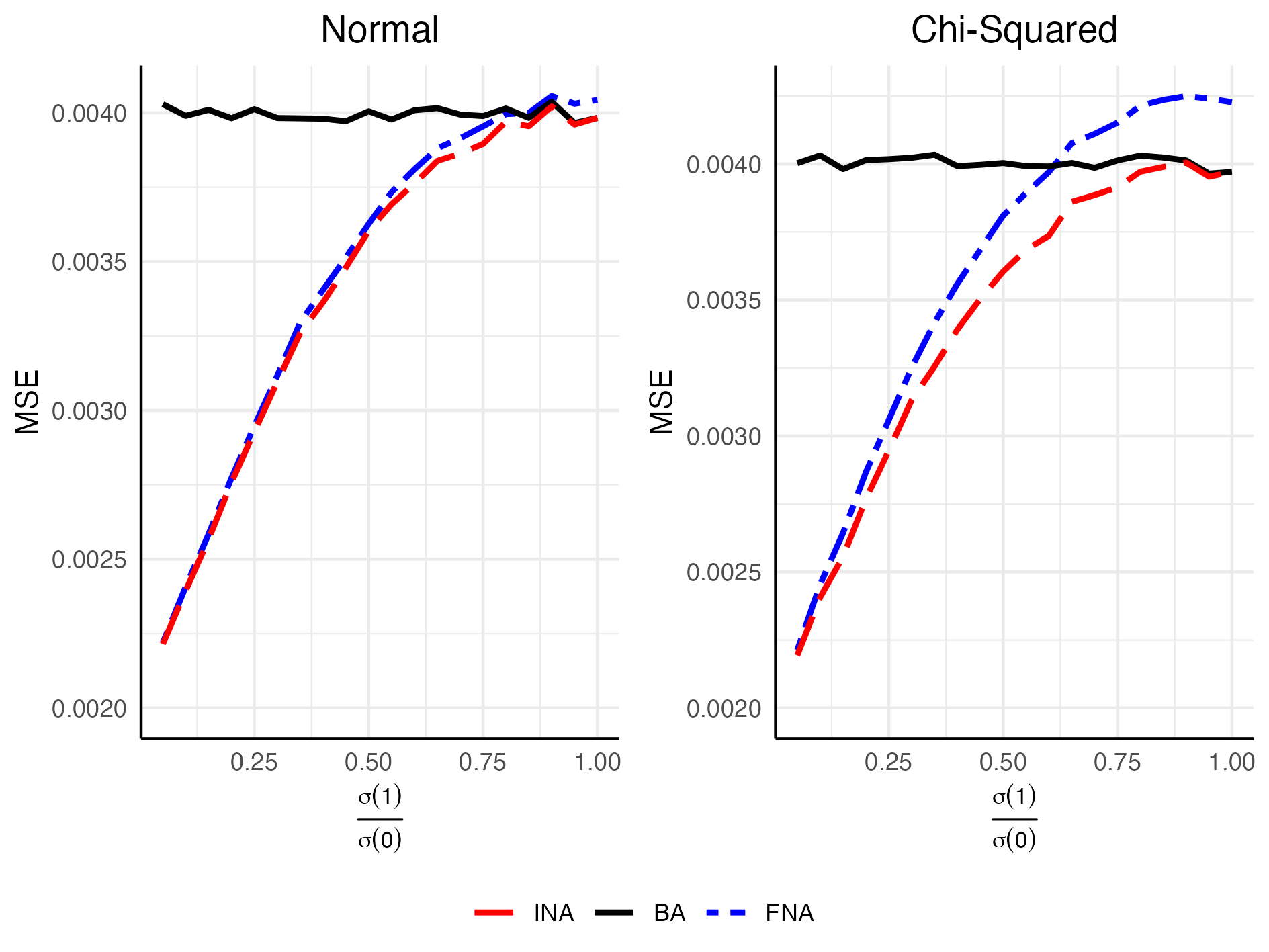}
\vspace{-5mm}
\caption{\small MSE under the Infeasible Neyman Allocation (INA), Balanced Allocation (BA) and Feasible Neyman Allocation (FNA) when $\varepsilon(1), \varepsilon(0)$ has the Normal distribution ($N(0,1)$) and the standardized Chi-Squared distribution ($\chi^2_1 - 1$).}
\vspace{5mm}
\includegraphics[width=1\linewidth]{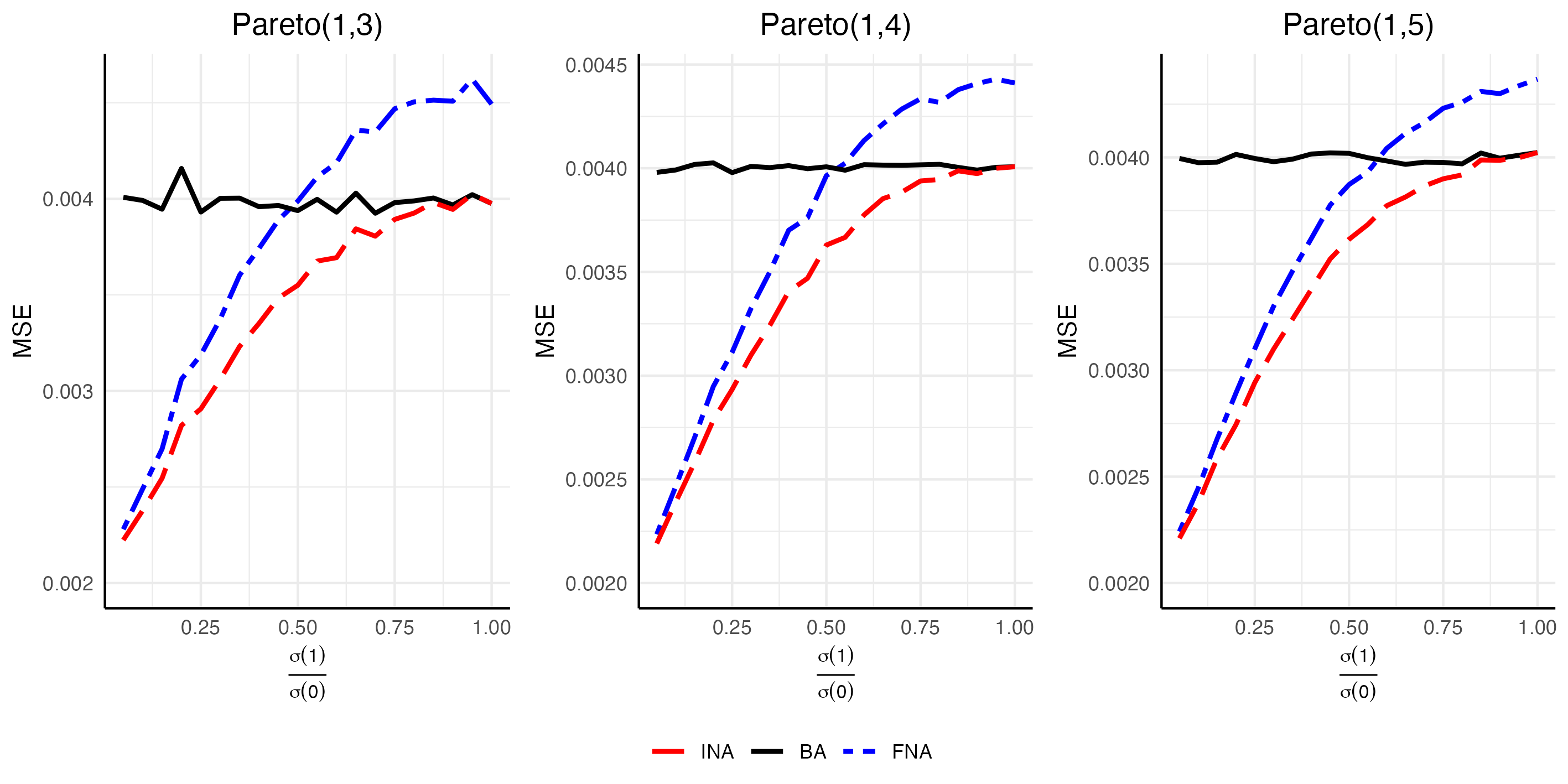}
\vspace{-5mm}
\caption{\small MSE under the Infeasible Neyman Allocation (INA), Balanced Allocation (BA) and Feasible Neyman Allocation (FNA) when $\varepsilon(1), \varepsilon(0)$ has the standardized Pareto($1,s$) distribution, with $s \in \{3,4,5\}$.}
\end{figure}

\newpage

\subsection{MSE of Various Solutions with Pareto Distribution}\label{appendix--sim_solutions}
\begin{figure}[h!]
\centering
\includegraphics[width=1\linewidth]{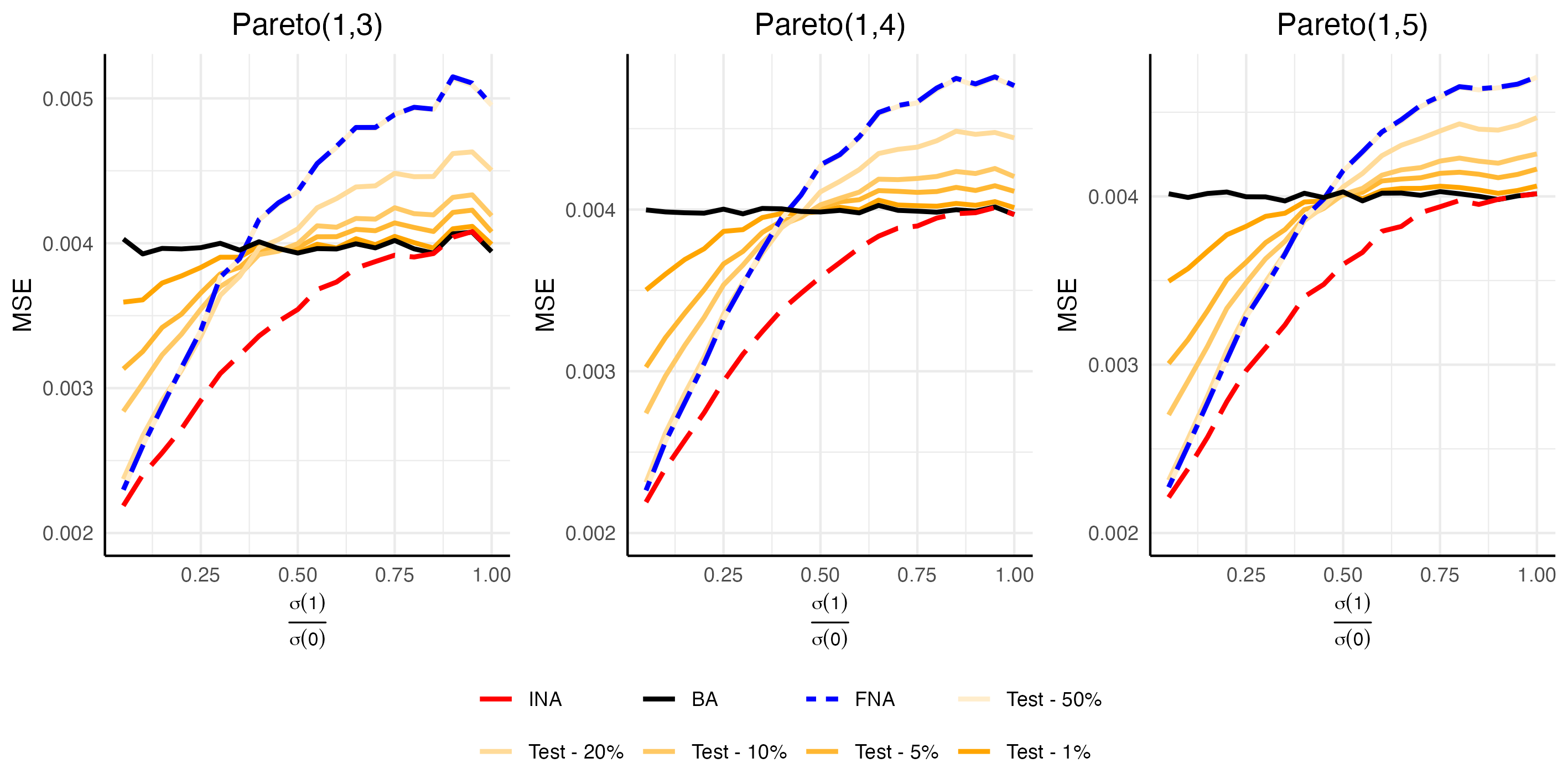}
\vspace{-5mm}
\caption{\small MSE under the Infeasible Neyman Allocation (INA), Balanced Allocation (BA) and Feasible Neyman Allocation (FNA) when $\varepsilon(1), \varepsilon(0)$ has the standardized Pareto($1,s$) distribution, with $s \in \{3,4,5\}$.}
\vspace{5mm}
\includegraphics[width=1\linewidth]{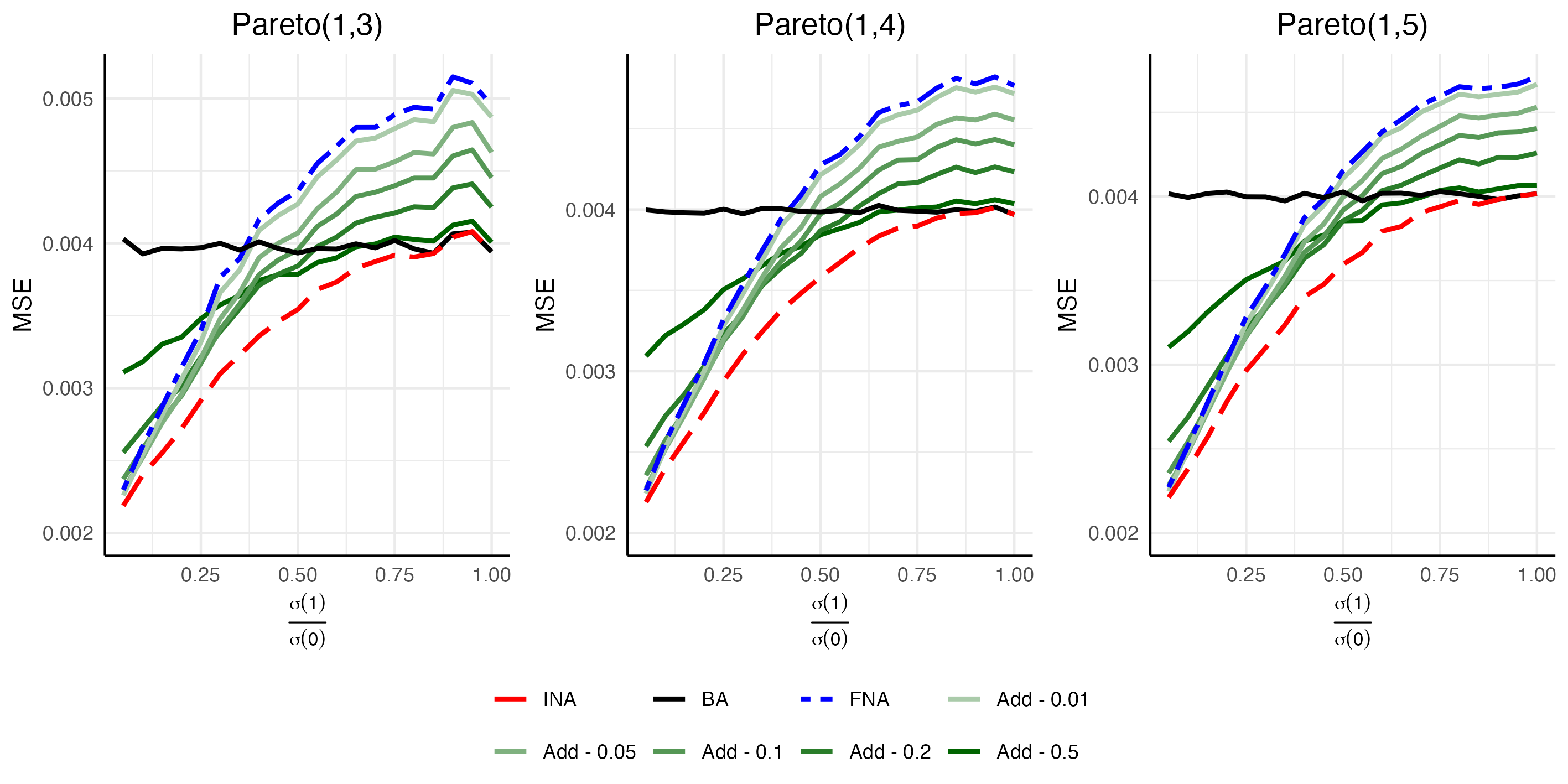}
\vspace{-5mm}
\caption{\small MSE under the Infeasible Neyman Allocation (INA), Balanced Allocation (BA) and Feasible Neyman Allocation (FNA) when $\varepsilon(1), \varepsilon(0)$ has the standardized Pareto($1,s$) distribution, with $s \in \{3,4,5\}$.}
\end{figure}

\begin{figure}[h]
\centering
\includegraphics[width=1\linewidth]{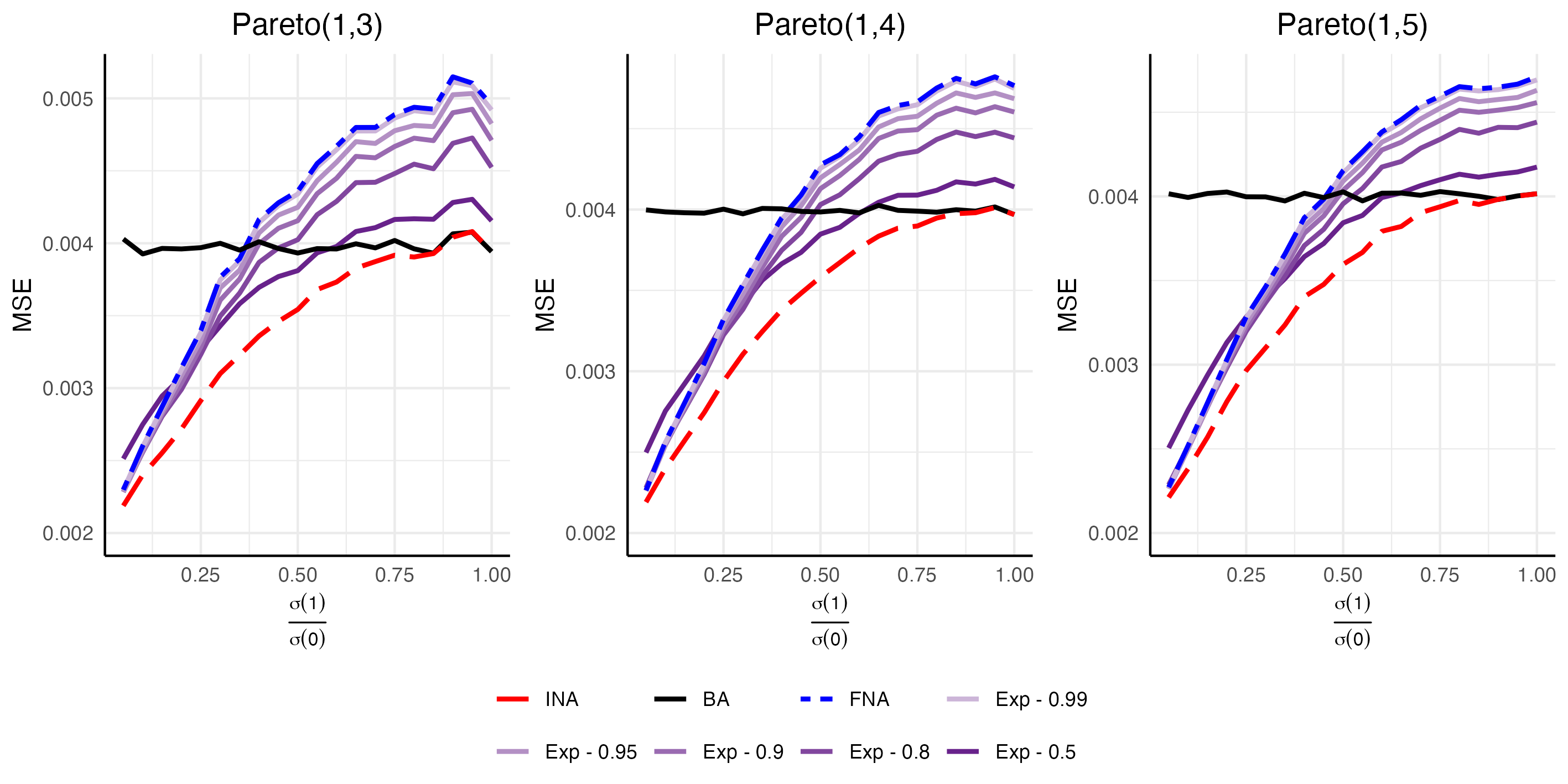}
\vspace{-5mm}
\caption{\small MSE under the Infeasible Neyman Allocation (INA), Balanced Allocation (BA) and Feasible Neyman Allocation (FNA) when $\varepsilon(1), \varepsilon(0)$ has the standardized Pareto($1,s$) distribution, with $s \in \{3,4,5\}$.}
\end{figure}

\begin{figure}[h]
\centering
\includegraphics[width=1\linewidth]{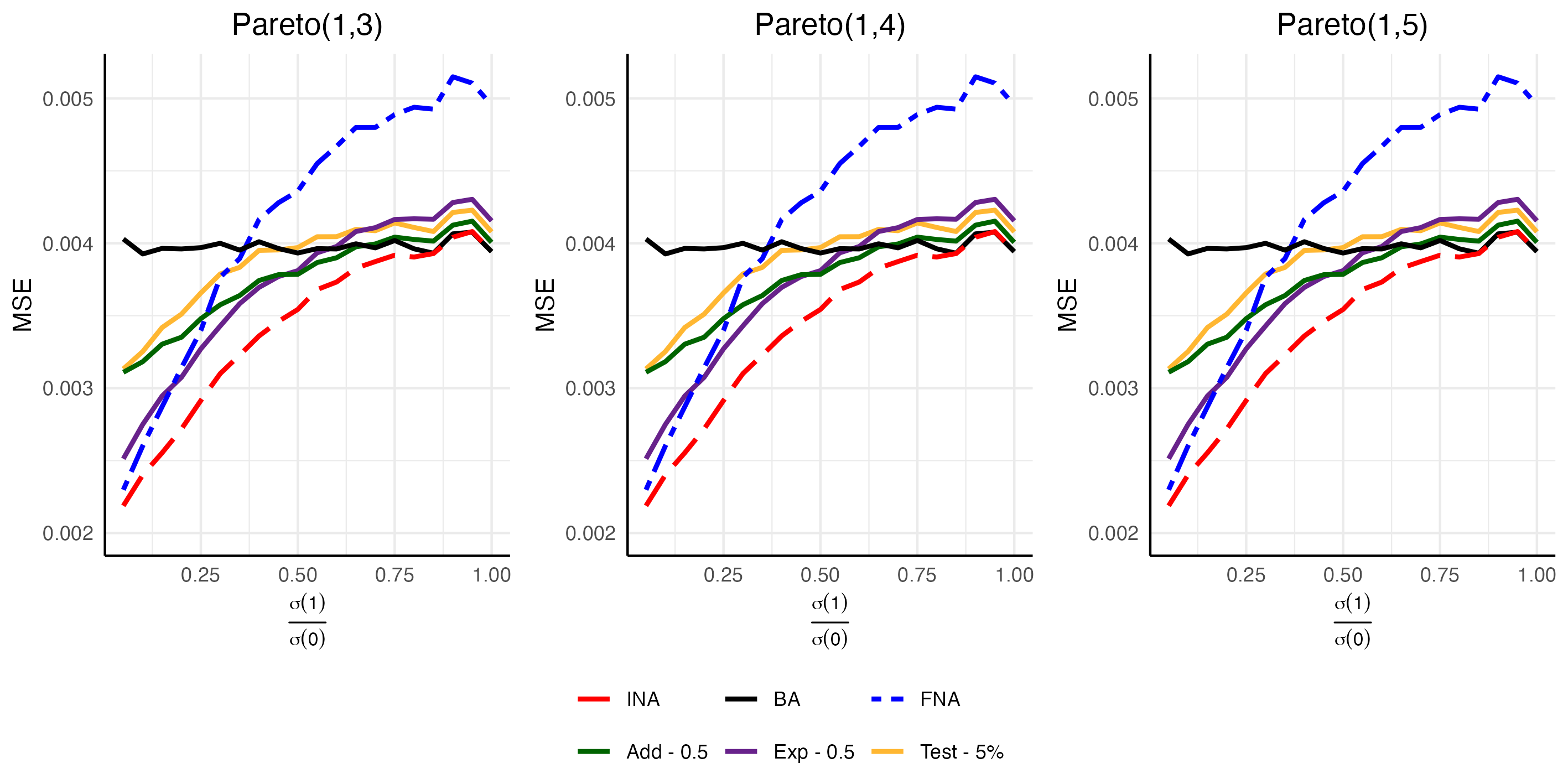}
\vspace{-5mm}
\caption{\small MSE under the Infeasible Neyman Allocation (INA), Balanced Allocation (BA) and Feasible Neyman Allocation (FNA) when $\varepsilon(1), \varepsilon(0)$ has the standardized Pareto($1,s$) distribution, with $s \in \{3,4,5\}$.}
\end{figure}

\end{document}